\documentclass[showpacs,preprintnumbers,amsmath,amssymb,twocolumn,
prd,a4paper,nofootinbib,superscriptaddress,showkeys]{revtex4-1} 

\usepackage{graphicx}
\usepackage{dcolumn}
\usepackage{bm}
\usepackage{amsmath}
\usepackage{amssymb}
\usepackage{dsfont}
\usepackage{amsfonts}
\usepackage[UKenglish]{babel}
\usepackage{hyperref}
\usepackage{nicefrac}
\usepackage{verbatim}
\usepackage{bbm}
\usepackage{color}

\usepackage{tabularx}

\newcommand{\bx}{\boldsymbol{x}}

\newcommand{\bp}{\boldsymbol{p}}
\newcommand{\bq}{\boldsymbol{q}}

\renewcommand{\vec}[1]{\mbox{\boldmath$#1$\unboldmath}}

\newcommand{\bigO}{\mathcal{O}}
\newcommand{\tr}[1]{\,\text{tr}\left[#1\right]}

\newcommand{\dthree}[1]{\textrm{d}^3#1\,}

\begin{document}

\title{Overlap Quark Propagator in Coulomb Gauge QCD and\\the Interrelation of Confinement and Chiral Symmetry Breaking}

\author{M.~Pak}
\email{markus.pak@uni-graz.at}
\affiliation{Institut f\"ur Physik, FB Theoretische Physik, Universit\"at Graz, Universit\"atsplatz 5,
8010 Graz, Austria}
\author{M.~Schr\"ock}
\email{mario.schroeck@roma3.infn.it}
\affiliation{Istituto Nazionale di Fisica Nucleare (INFN), Sezione di Roma Tre, 00146 Rome, Italy}

\begin{abstract}
The chirally symmetric overlap quark propagator is explored in Coulomb gauge for the first time. This gauge is especially well suited for studying the interrelation between confinement and chiral symmetry breaking, since confinement can be attributed to the infrared divergence of the vector dressing function of the inverse quark propagator. Using quenched gauge field configurations on a $20^4$ lattice, the dressing functions of the quark propagator as well as the dynamical quark mass function are evaluated, also in the chiral limit. Chiral symmetry is artificially restored by removing the low mode contribution from the quark propagator. After removing enough low-lying modes, the dynamical quark mass function approaches the current quark mass in the whole momentum region and goes to zero in the chiral limit. However, the vector dressing function is unaffected by the low-mode removal. It follows that the quark energy dispersion is still infrared divergent and quarks in such a chiral symmetry restored phase are still confined within a hadron. 
\end{abstract}
\maketitle

\section{Introduction}
Spontaneous chiral symmetry breaking and
confinement are the two main phenomena of low energy QCD. Chiral symmetry breaking goes
in hand with dynamical mass generation, from which the low-lying hadron spectrum can be
understood. However, the underlying mechanism of dynamical mass generation and the interplay
between chiral symmetry breaking and confinement are still unresolved questions. 

The behavior of propagators allows to shed light on these unresolved issues.
Coulomb gauge, where a confining static quark potential arises
naturally, is especially well suited for this purpose. From
continuum mean-field studies using the so-called variational
approach or Dyson-Schwinger equations it has been
conjectured that confinement in this gauge reflects itself via
infrared divergent quark propagator dressing functions, Refs.~\cite{Adler:1984ri,
Alkofer:1988tc, Wagenbrunn:2007ie}, which, however, give an infrared constant dynamical
mass function, interpreted as the constituent mass of the quark. Only recently there has been
a first attempt to transport this picture to full QCD using the
lattice approach, Refs.~\cite{Burgio:2012ph, Burgio:2013mx}, giving hints that the quark self-energy is 
indeed diverging in the low-momentum region.
However, a quark propagator analysis with chirally symmetric
lattice fermions is still lacking in Coulomb gauge. We attempt
to close this gap. Overlap Dirac propagator studies in Landau
gauge can be found in Refs.~\cite{Bonnet:2002ih,
Zhang:2003faa,Zhang:2004gv,Trewartha:2013qga}.

Continuum studies in Coulomb gauge are, in principle, appealing, for 
instance to explore the region of finite chemical potential \cite{Guo:2009ma} and to systematically
analyze the excited meson spectrum \cite{Pak:2013cpa}. However, although some progress has been made in recent years
\cite{Pak:2011wu, Pak:2013uba}, continuum quark propagator studies
in Coulomb gauge are still ansatz-dependent and far from being
conclusive. Therefore, input from the lattice is highly needed.

One of the main motivations
of this work is to explore, on the level of Green's functions, what
happens to the confinement properties of quarks, when chiral
symmetry is artificially restored by removing the low-lying eigenmodes from 
the Dirac operator in the valence quark sector. In recent years the effect of
such an artificial symmetry restoration on the hadron spectrum
has been analyzed, see Refs.~\cite{Lang:2011vw, Glozman:2012fj,
Denissenya:2014poa}.  From the exponential decay signal of hadron correlators
 it could indirectly be deduced that  the confinement properties stay intact. 
Here we want to give a more direct picture of the confinement
issue by analyzing the low-mode truncated quark propagator in Coulomb gauge. 
The central question is how do the dressing functions change their shape when 
chiral symmetry is artifically restored and whether an IR-diverging dispersion relation
is still possible.
A first study of the low-mode truncated quark propagator has been made in
Ref.~\cite{Schrock:2011hq}, however, where Landau gauge has been used, and
a lattice fermion discretization, which only approximately fulfills the
Ginsparg-Wilson equation.

Having a chirally symmetric
lattice Dirac operator at hand, is central for our purpose. 
We use the overlap Dirac operator, Refs.~\cite{Neuberger:1997fp,
Neuberger:1998wv}, which is an explicit solution
to the Ginsparg-Wilson equation. Besides its nice properties, 
like the occurence of exact zero modes, it admits a
clear prescription to extract the dressing functions of the quark
propagator. After identifying the overlap lattice momenta, no
further tree-level corrections or improvements have to be
performed, which could lead to ambiguities in the result. On the other hand, 
since overlap fermions are numerically very
costly, we are restricted to rather small lattices. Nevertheless, we can enter
the small momentum region of the dynamical mass function and evaluate the low energy
chiral properties of QCD. 

The organization of the article is as
follows: in Chapter~\ref{Overlap-Intro} the overlap Dirac operator
is defined, the free case is discussed and the overlap lattice
momenta are derived. In Chapter~\ref{Lattice-Setup} the lattice setup, the numerical implementation
of the overlap Dirac operator and the gauge fixing procedure are described. In Chapter~\ref{Chap-Coulomb-Gauge} the Coulomb gauge quark propagator is
introduced and the numerical results for the dressing functions are presented. In Chapter~\ref{Chap-Low-Mode} the Dirac low mode removal is discussed, its relation to confinement is clarified
and the dressing functions for different truncation steps are shown. In Chapter~\ref{Chap-Summary} the main results of this work are summarized and a short conclusion and outlook are given. 

\section{Overlap Dirac Operator and Free Propagator}
\label{Overlap-Intro}
The massless overlap Dirac operator is defined as 
\begin{align}
\label{Overlap-Def}
 D(0) = \rho \left( \mathds{1} + \gamma_5 \mbox{sign}\left[H_{\textsc{W}}(-\rho)\right] \right) \; ,
\end{align}
with $H_{\textsc{W}}(-\rho) = a \gamma_5 D_{\textsc{W}}(-\rho)$ the (dimensionless) Hermitian Wilson-Dirac operator and $\rho$ a negative mass parameter.
The operator (\ref{Overlap-Def}) fullfills the Ginsparg-Wilson equation \cite{Ginsparg:1981bj}
\begin{align}
\label{GW-eq}
 D(0) \gamma_5 + \gamma_5 D(0) = \frac{1}{\rho} D(0) \gamma_5 D(0) \; ,
\end{align}
and therefore represents exact chiral symmetry on the lattice, see Refs.~\cite{Hasenfratz:1998ri, Hasenfratz:1998jp, Luscher:1998pqa}.
The eigenvalues of $D(0)$ lie on a circle in the complex plane with radius $\rho$.
The free massless overlap quark propagator in momentum space, denoted as $S^{(0)}(p)$, can be decomposed as
\begin{align}
S^{(0)}(p) = - i  \gamma_{\mu} \mathcal{C}_{\mu}(p) + \frac{1}{2\rho} \; ,
\end{align}
with
\begin{align}
 \mathcal{C}_{\mu}(p) = \frac{1}{2 \rho} \frac{k_{\mu}}{\sqrt{k^2_{\mu}+A^2}+A} \; , \qquad A = \frac{1}{2} \hat{k}^2_{\mu} - a \rho \; ,
\end{align}
and the momenta $k_{\mu} = \sin(p_{\mu} a),  \hat{k}_{\mu} = 2 \sin(p_{\mu} a/2) $.
In contrast to its continuum counterpart, the free massless quark propagator has an additional term $1/(2\rho)$, which enters with the unit matrix
in Dirac space. It is allowed in the Ginsparg-Wilson equation (\ref{GW-eq}).

According to Ref.~\cite{Bonnet:2002ih}, we redefine the massless overlap quark propagator as
\begin{align}
 \widetilde{S} = S - \frac{1}{2\rho} \; ,
\end{align}
which corresponds to imposing continuum chiral symmetry for overlap fermions: $\{\widetilde{S},\gamma_5 \} = 0$. The free massless lattice quark propagator
then has the same Dirac structure as the free continuum propagator. 
No additional improvement techniques have to be imposed, which could introduce a small arbitrariness, depending on the particular procedure, see Ref.~\cite{Bonnet:2002ih}. This is a great advantage of choosing overlap fermions.

\begin{figure}[t]
 \centering
 \includegraphics[angle=0,width=.98\linewidth]{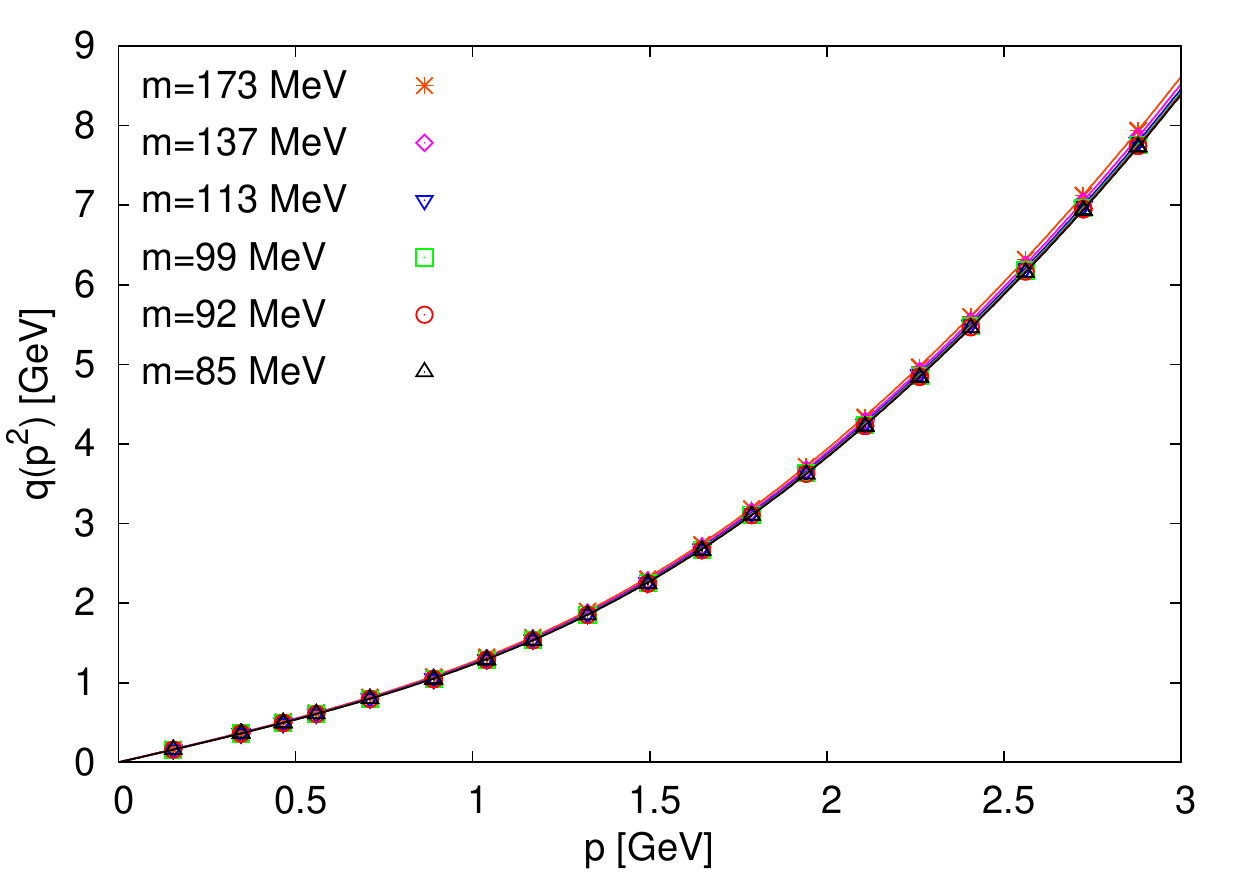}
 \caption[Lattice momenta and mass]{\sl
 Overlap lattice momentum $q(p)$ for six different masses in units of MeV ($a=0.2$ fm).}
 \label{free-latt}
 \end{figure}

\begin{figure}[t]
 \centering
 \includegraphics[angle=0,width=.98\linewidth]{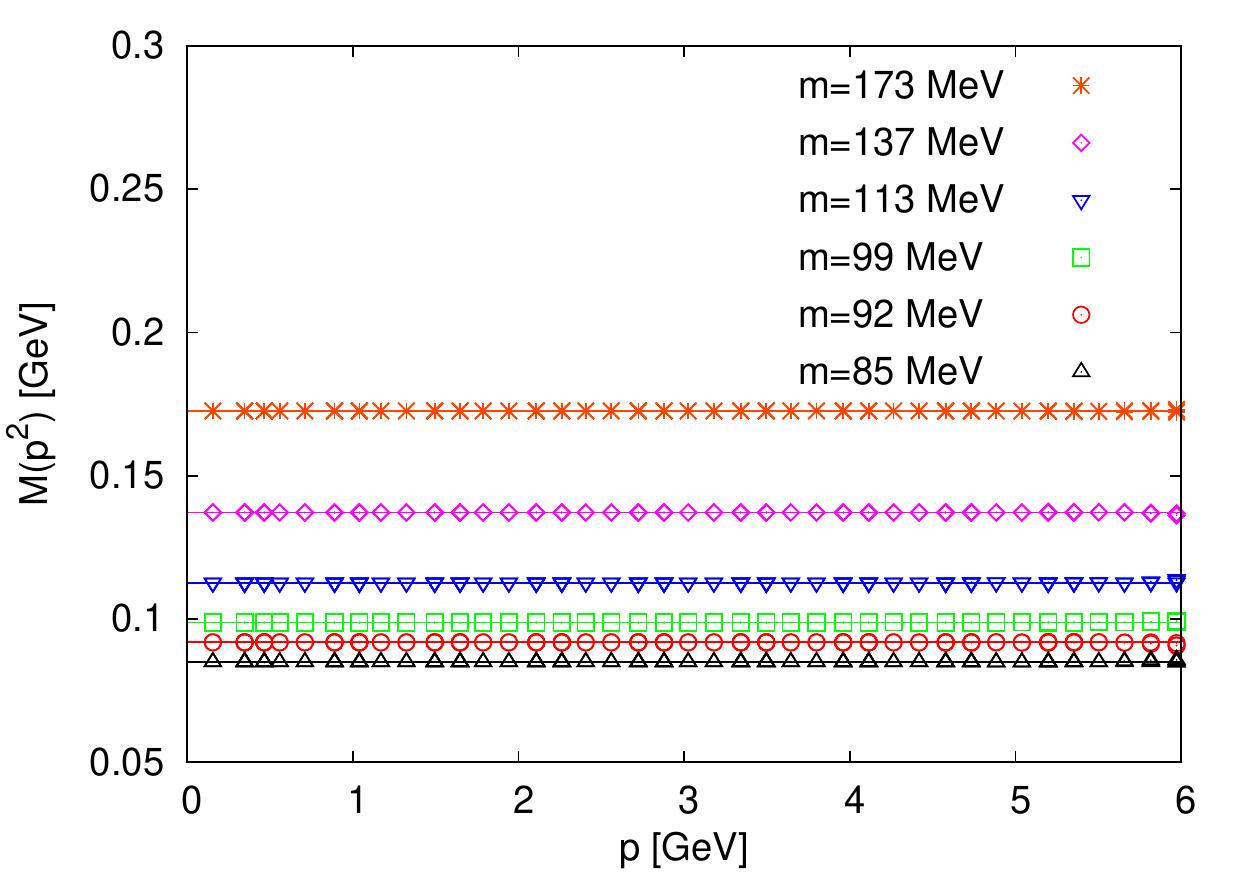}
 \caption[Lattice momenta and mass]{\sl
 Current quark mass $m$ for free overlap fermions.}
 \label{free-latt2}
 \end{figure}

The massive overlap
Dirac operator follows from Eq.~(\ref{Overlap-Def}) via
\begin{align}
 D(m_0) = \left( 1 - \frac{m_0}{2\rho} \right) D(0) + m_0 \; ,
\end{align}
with $m_0$ the quark mass parameter. From the massive free (inverse) overlap quark propagator, which
has the structure
\begin{align}
 \left(\widetilde{S}^{(0)}\right)^{-1}(p) = i \gamma_{\mu} q_{\mu} + \mathds{1} m \; ,
\end{align}
we identify the overlap lattice momenta $q_{\mu}$ and the current quark mass $m$
\begin{align}
\label{qmu}
 q_{\mu} = \frac{4\rho^2}{(2\rho -m_0)}\frac{k_{\mu}\left(\sqrt{k^2_{\mu}+A^2}+A \right)}{k_{\mu}^2}\; , \quad m = \frac{m_0}{1-\frac{m_0}{2\rho}} \; .
\end{align}
As a first test of our code we compare the numerically evaluated free overlap quark propagator with the analytic formulas (\ref{qmu}), see Figs.~\ref{free-latt}, \ref{free-latt2}. 
A similar plot of the overlap lattice momenta $q_{\mu}(p)$ can be found in Ref.~\cite{Bonnet:2002ih}.
Throughout this work we perform a cylinder cut \cite{Skullerud:2000un} on all data, that is we evaluate the numerical functions only within one unit of spatial momentum from the diagonal of four-momentum space.

 \section{Setup of the Calculation}
 \label{Lattice-Setup}
 
 \subsection{Gauge field configurations}
We use quenched  L\"uscher-Weisz \cite{Luscher:1984xn} gauge field configurations which we generated with the Chroma software 
package \cite{Edwards:2004sx} and QDP-JIT \cite{Winter:2014npa, Winter:2014dka}.
We adopted an inverse coupling value $\beta_1 = 7.552$ which corresponds to a lattice spacing $a=0.2$ fm \cite{Gattringer:2001jf}.\footnote{Both, the L\"uscher-Weisz gauge action and the overlap fermionic action are improved, to orders $\bigO(a^2)$ and $\bigO(a)$, respectively, which allow the usage of a rather large lattice spacing. Moreover, this study focuses on the infrared behaviour of QCD, where the ultraviolet cutoff is of minor importance.}
The size of the lattice is $N_\textsc{S} \times N_{\textsc{T}} = 20^3 \times 20$ and the value of the average plaquette is $0.5767$. 

We use an ensemble of $96$ well separated configurations and evaluate the quark propagator for six current quark masses chosen at $m=85, 92, 99, 113, 137, 173$ MeV. 
The quark propagators are obtained by inverting the overlap Dirac matrix for each configuration on point-sources. 
Subsequently, the quark propagators are transformed to momentum space after which the extraction of the dressing functions 
is performed according to the description in \cite{Skullerud:1999gv, Burgio:2012ph}.

\subsection{Gauge fixing}
The continuum Coulomb gauge condition,
\begin{equation}
\partial_iA_i = 0,\quad i=1, 2, 3,
\end{equation}
can be realized on the lattice by maximizing the gauge functional
\begin{equation}
F_g[U] = \Re\sum_{ i, x} \mathrm{tr}\big[ g(x)\, U_i(x) \, g(x+\hat i)^\dagger\big]
\label{eq:F}
\end{equation}
with respect to gauge transformations $g(x)\in\mathrm{SU}(3)$.
The gauge links maximizing Eq.~(\ref{eq:F}) satisfy
the discretized Coulomb gauge condition
\begin{equation}
\Delta^g(x)\equiv \sum_i\left(A_i^g(x)-A_i^g(x-\hat i) \right) =0\,,
\end{equation}
in which the linear definition of the gauge fields is used,
\begin{equation}
A_i(x)\equiv \left[\frac{U_i(x)-U_i(x)^\dagger}{2iag_0}\right]_{\textrm{traceless}}
\end{equation}
where $g_0$ ist the coupling constant.

We maximize $F_g[U]$ with the cuLGT code \cite{Schrock:2012fj} and the overrelaxation algorithm 
\cite{Mandula:1990vs}, which in the case of Coulomb gauge can be applied 
to each time-slice independently.

A measure of the quality of the gauge fixing 
is the average $L_2$-norm of the gauge fixing violation $\Delta^g \neq 0$ 
\begin{equation}
\theta(x_4) \equiv \frac{1}{N_S^3N_c}\sum_{\vec x}\tr{\Delta^g(\vec{x},x_4)
\Delta^g(\vec{x},x_4)^\dagger}\, ,
\end{equation}
where the sum runs over all spatial sites $\vec x$ within one time-slice and $N_S$ is the 
number of spatial lattice sites.
We have chosen $\theta< 10^{-12}$ as the stopping criterion for the 
overrelaxation algorithm.

In Coulomb gauge, maximizing Eq.~(\ref{eq:F}) leaves the temporal links 
$U_4(x)$ unfixed, i.e., a residual gauge freedom with respect to 
space independent gauge transformations $g(x_4)$ is left. 
One possible choice to fix the residual gauge in the 
continuum is to require
\begin{equation}
\partial_4 \int \dthree{x} A_4(x) =0\,.
\label{eq:tgauge}
\end{equation}
We will use the lattice version of Eq.~(\ref{eq:tgauge}),
the so-called integrated Polyakov gauge \cite{Burgio:2008jr}.

\subsection{Overlap Dirac operator}
The (valence) overlap Dirac operator is implemented in the following way: the low-lying part of $\mbox{sign}[H_{\textsc{W}}]$ is calculated via the spectral decomposition.
For the high-lying part the Chebyshev approximation is used, see Ref.~\cite{Gattringer:2010zz}. Eigenvalues and -vectors are extracted with the ARPACK routines, Ref.~\cite{Arpack}.
For the inversion of the overlap Dirac operator the biconjugate gradient stabilized multimass solver is used, see Ref.~\cite{Jegerlehner:1996pm}.
The Wilson-Dirac mass parameter $\rho$ is set to $1.6$.

In order to reduce the cost for the overlap Dirac operator, three sweeps of stout smearing \cite{Morningstar:2003gk}
are applied on the gauge field configurations. This procedure is especially important when computing the low-lying eigenmodes of $D$. 
After three sweeps of stout smearing the configurations are smooth enough so that the low modes of the
Hermitian Wilson-Dirac operator are suppressed, see Ref.~\cite{Neuberger:1999pz} for details. This results in a 
speed-up of the overlap construction by a factor of three.

The propagators are computed on unsmeared and stout smeared configurations.
The low-mode truncated propagators, i.e. where the low-lying eigenvalues are removed 
from the propagators, are computed on smeared configurations only, since
for the evaluation of the lowest overlap eigenvalues smearing is essential to reduce the
computational cost. In Chapter \ref{stout-compare} we analyze 
the effect of the smearing procedure on the quark propagator dressing functions. 

\section{Non-Perturbative Quark Propagator in Coulomb Gauge}
\label{Chap-Coulomb-Gauge}

\subsection{Introduction}

In Coulomb gauge, due to the breaking of Lorentz invariance, four independent scalar dressing functions
for the (Euclidean) quark propagator occur:
\begin{align}
\label{prop-parameterization}
 S^{-1}(\bp, p_4) = i &\gamma_i p_i A_{\textsc{s}}(\bp,p_4) + i \gamma_4 p_4 A_{\textsc{t}}(\bp,p_4)  \\ 
+ & \gamma_4 p_4 \gamma_i p_i A_{\textsc{d}}(\bp,p_4) + \mathds{1} B(\bp,p_4) \nonumber \; ,
\end{align}
with $A_{\textsc{s}}, A_{\textsc{t}}, A_{\textsc{d}}, B $ referring to \textit{spatial}, \textit{temporal}, \textit{mixed} and \textit{scalar} dressing functions \cite{Popovici:2008ty}, respectively.
Here $\bp$ denotes 3-momentum. 
For free fermions in the (Euclidean) continuum the dressing functions are $A_{\textsc{s}} = A_{\textsc{t}}=1, B = m_0 $ and  $A_{\textsc{d}}=0$. 

From a recent study in (lattice) Coulomb gauge, Ref.~\cite{Burgio:2012ph}, it has been shown that the dressing functions in Eq.~(\ref{prop-parameterization}) are independent
of the $p_4$-component, which is also supported by our results with overlap fermions. 
Moreover, we also find that the mixed component $A_{\textsc{d}}(\bp,p_4)$ vanishes.
For the smallest momenta, however, the data suffer from large statistical fluctuations. 
To increase statistics, we average the dressing functions over the lattice cubic symmetries, the parity symmetry of the quark propagator and finally over the $p_4$-component.

From mean-field studies in continuum Coulomb gauge it is suspected that, due to the presence of the non-Abelian color-Coulomb potential, the scalar and 
vector dressing functions diverge for $|\bp| \rightarrow 0$, see Refs.~\cite{Alkofer:1988tc, Wagenbrunn:2007ie}. 
 The dynamical quark mass $M(\bp)$, which is defined as
\begin{align}
\label{dynamical-mass}
 M(\bp) = \frac{B(\bp)}{A_{\textsc{s}}(\bp)} \; , 
\end{align}
however, reaches a constant value for $|\bp| \rightarrow 0$, identified as the constituent quark mass. The infrared divergencies of $B(\bp)$ and 
$A_{\textsc{s}}(\bp)$ cancel each other, so that $M(\bp)$ is infrared finite, see Refs.~\cite{Watson:2011kv, Watson:2012ht} for details. 
It is this divergence property of scalar and vector dressing functions that makes Coulomb gauge special, since it can be related to
the confinement properties of quarks, as we will show in the next section.

We stress that in the continuum non-linear integral equations (Dyson-Schwinger equations) an infrared regulator has to be introduced to make the integrals, due to the non-Abelian color-Coulomb potential, well-defined.   
The divergence of the dressing functions appears for vanishing infrared regulator. For a finite lattice the dressing functions then should approach a constant value for $|\bp|=0$.  

On the lattice, the regularized quark propagator is extracted which depends on the lattice cutoff $a$. The regularized quark propagator
$S_L(\bp;a)$ can be renormalized at the renormalization point $\xi$ 
with the momentum independent quark wave-function renormalization constant $Z_2(\xi;a)$,
\begin{equation}
	S_L(\bp;a) = Z_2(\xi;a) S(\xi;\bp).
\end{equation}
While the mass function $M(\bp)$ is independent of the renormalization point $\xi$, the wave-function renormalization function 
\begin{equation}\label{Z-chi}
	Z_{\xi}(\bp)\equiv 1/A_{\textsc{s}}(\bp)
\end{equation}
differs at different scales but can be related
by multiplication with a constant.

\subsection{Numerical results}

Here we discuss our results for the dressing functions of the inverse quark propagator $S^{-1}(\bp)$,  Eq.~(\ref{prop-parameterization})
as well as the dynamical quark mass $M(\bp)$, Eq.~(\ref{dynamical-mass}), and the wave-function renormalization function $Z_{\xi}(\bp)$, Eq.~(\ref{Z-chi}).

\begin{figure}[t]
 \centering  
 \includegraphics[angle=0,width=.98\linewidth]{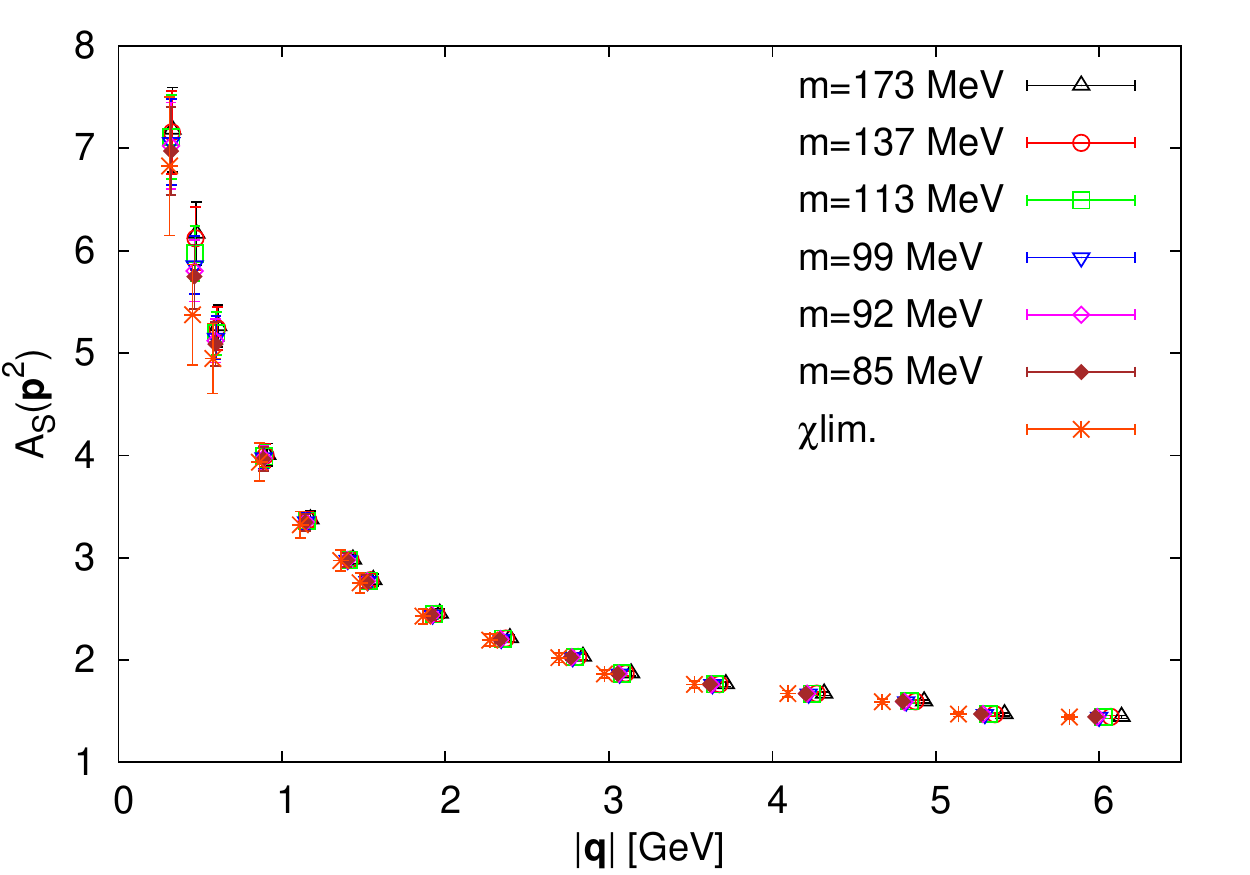}
 \caption[Spatial component for full case]{\sl
  Spatial component $A_{\textsc{s}}(\bp)$ for several quark masses and in the chiral limit.}
 \label{spatial-component}
 \end{figure}

\begin{figure}[t]
 \centering
 \includegraphics[angle=0,width=.98\linewidth]{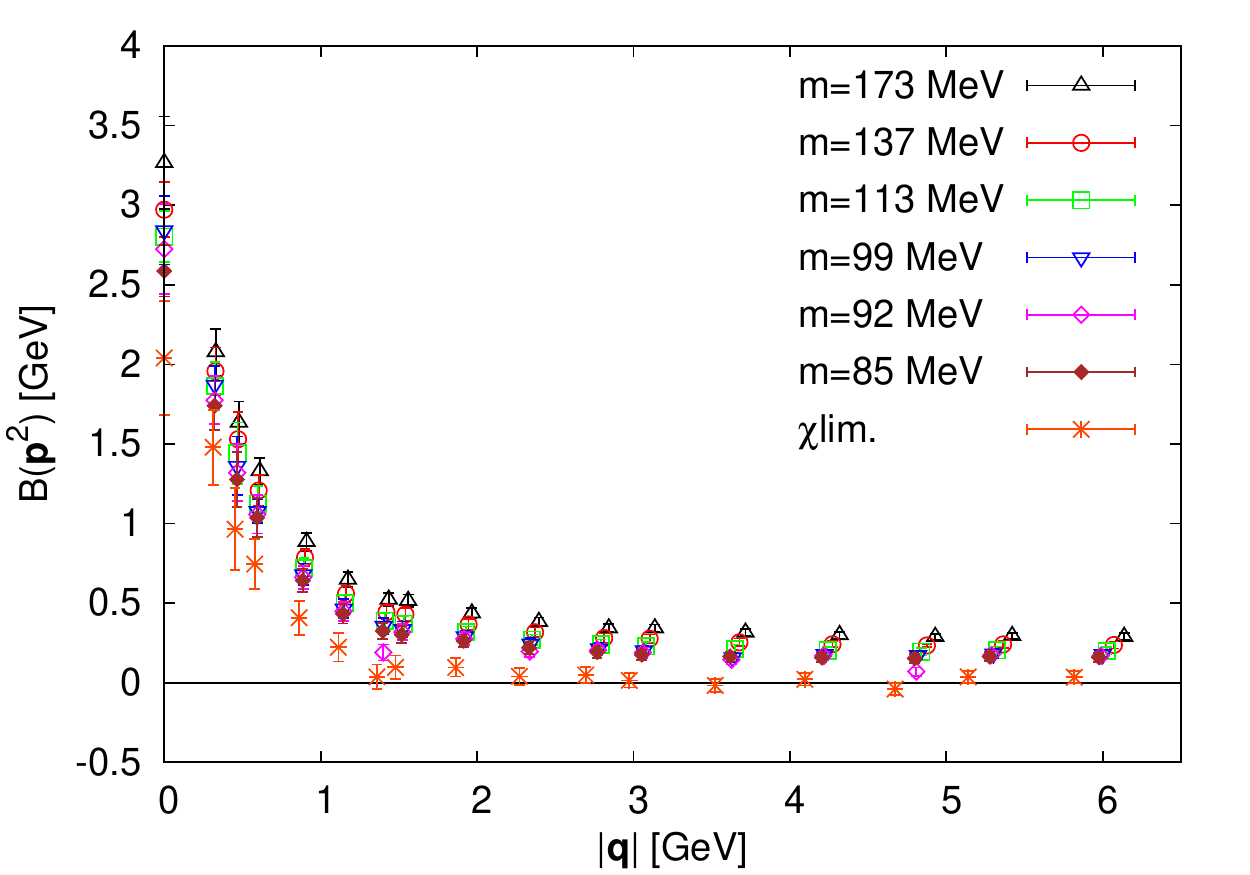}
 \caption[Lattice momenta and mass]{\sl
  Scalar component $B(\bp)$ for several quark masses and in the chiral limit.}
 \label{scalar-component}
 \end{figure}

\begin{figure}[t]
 \centering
 \includegraphics[angle=0,width=.98\linewidth]{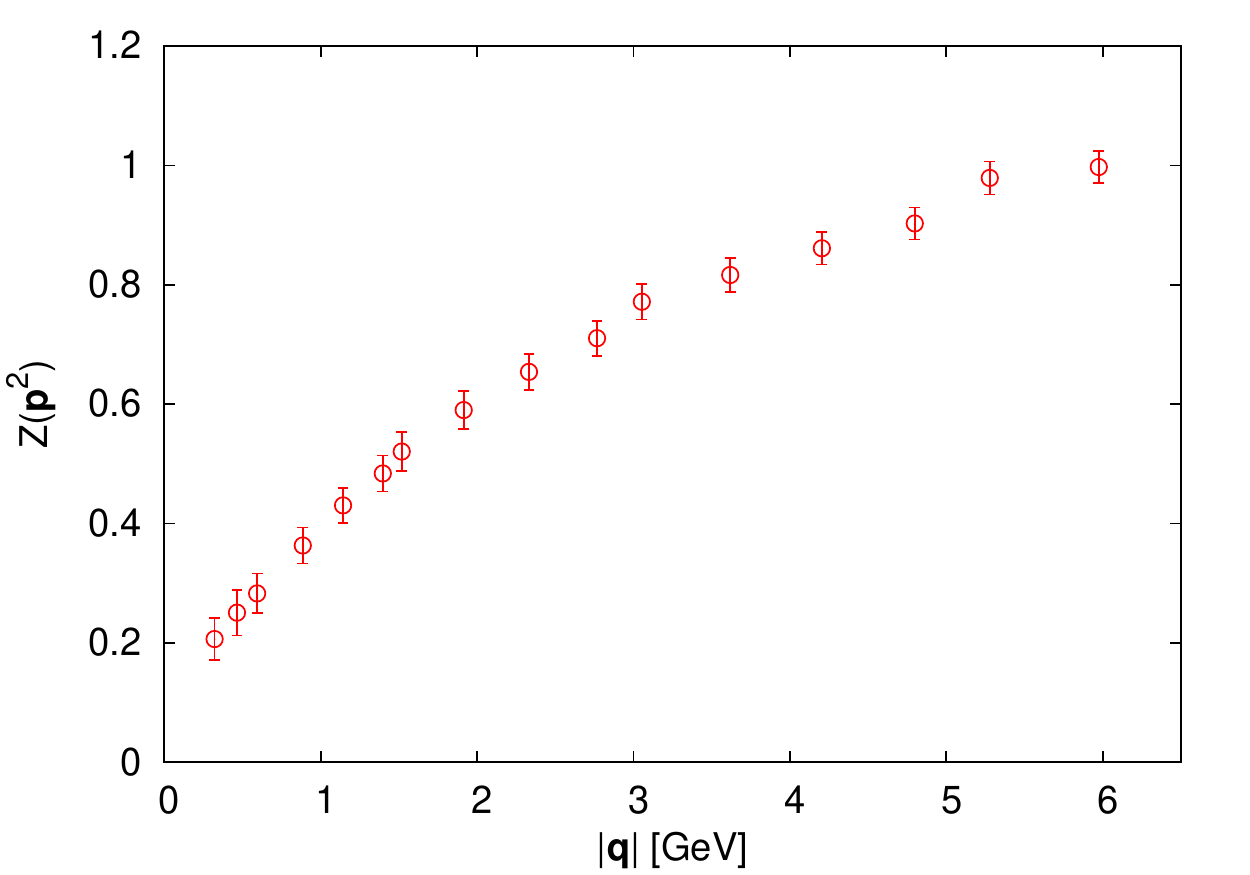}
 \caption[Lattice momenta and mass]{\sl
  $Z_{\xi}(\bp)$ renormalized at $\xi= 6$ GeV for smallest quark mass $m=0.085$ GeV.}
 \label{renormalization-func}
 \end{figure}

Most interesting are the spatial and scalar dressing functions,  $A_{\textsc{s}}(\bp)$ and $B(\bp)$. 
For large momenta $A_{\textsc{s}}(\bp)$ goes to unity and $B(\bp)$ approaches the current quark mass consistent with asymptotic freedom. 
Around one GeV both dressing functions increase for small momenta, which goes in hand with spontanteous symmetry breaking and 
dynamical mass generation, see Figs.~\ref{spatial-component},  
\ref{scalar-component}. We also show a simple linear extrapolation to the chiral limit. For massless quarks $B(\bp)$ 
goes to zero for $|\bp| \rightarrow \infty$.
We note that for zero momentum $B(\bp)$ reaches a finite value which is most likely a 
finite size effect. The vector dressing function $A_{\textsc{s}}(\bp)$ is not accessible at zero momentum. 
For the  scalar dressing function $B(\bp)$ larger current quark masses lead to larger values
for small momenta. The spatial dressing function $A_{\textsc{s}}(\bp)$ only has a mild mass dependence. All
values for the different masses lie within error bars. 
In Fig.~\ref{renormalization-func} the renormalization function $Z_{\xi}(\bp)$, renormalized at $\xi=6$ GeV, is shown. 
To verify that $A_{\textsc{s}}(\bp)$ is indeed IR diverging, or alternatively that $Z_{\xi}(\bp)$ goes to zero for $|\bp|\rightarrow 0$, 
the infinite volume limit has to be taken. This is
left to a further investigation. 

\begin{figure}[t]
 \centering
 \includegraphics[angle=0,width=.98\linewidth]{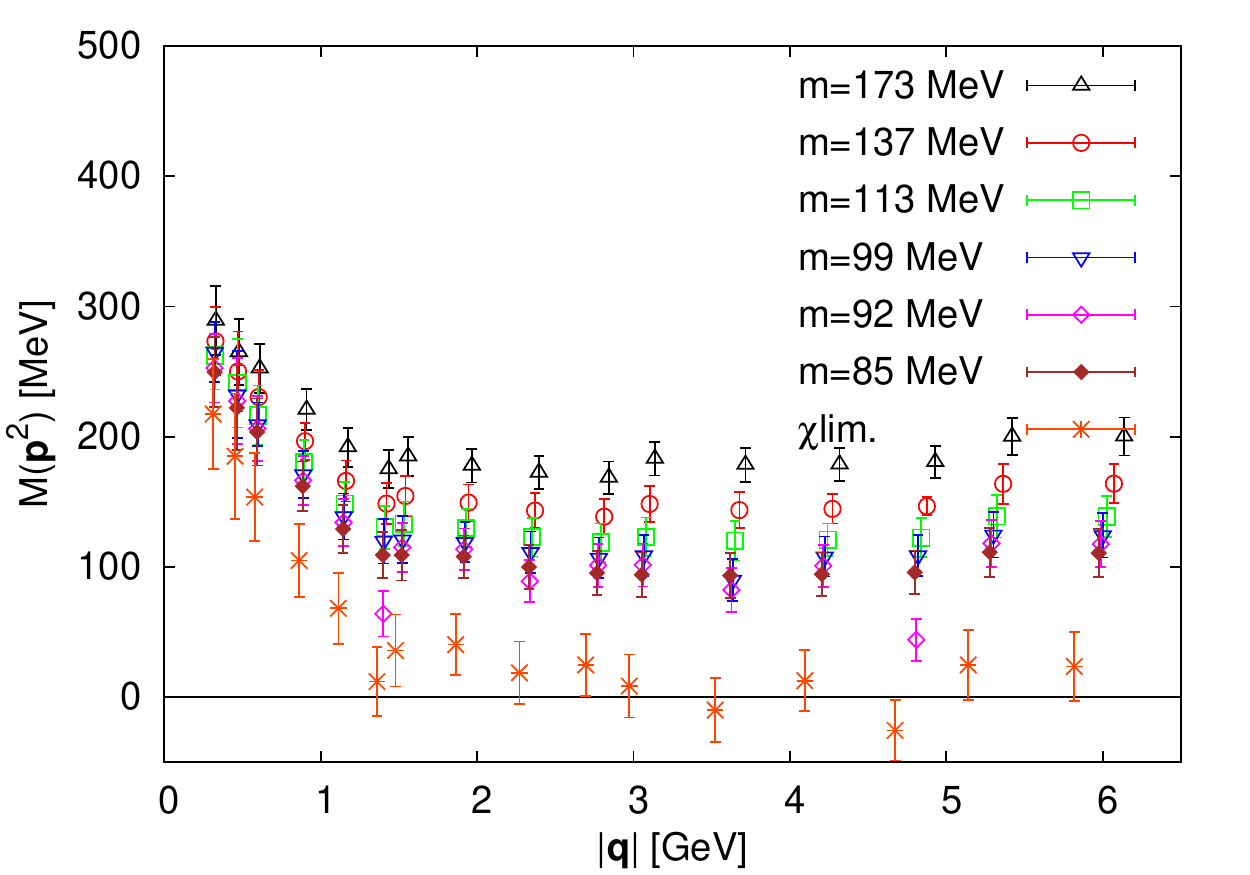}
 \caption[Lattice momenta and mass]{\sl
  Dynamical mass function $M(\bp)$ for several quark masses and in the chiral limit.}
 \label{dynamical-mass-plot}
 \end{figure}

The dynamical quark mass function $M(\bp)$ is shown in Fig.~\ref{dynamical-mass-plot}. 
For large momenta the mass function should approach the current quark mass $m$, which is, due to the rather coarse lattice, only approximately seen in our results. In the chiral limit it goes to zero, which is, within error bars, also approximately seen in our results. 
Between one and two GeV the function starts to increase 
and acquires large values for small momenta. The smaller the current quark mass, the more dynamical quark mass is generated. 
This means that for larger current quark masses the amount of IR mass generation is smaller. This becomes
obvious by comparing largest and smallest current quark masses $m=173$ MeV and $m=85$ MeV, respectively. At the
smallest accessible momentum their values are split only by around $50$ MeV. This behaviour reflects itself
in the strong infrared dynamical mass generation for chiral quarks. 

Due to the relatively large lattice we reach IR momenta down to around 300 MeV which
is, compared to former Coulomb gauge studies, quite an improvement. At $|\bq| \approx 300$ MeV the dynamical quark mass 
is already at 210 MeV for massless quarks. It is thus very likely that a constituent quark mass around 300 MeV is reached.

We also note that the dynamical quark mass $M(\bp)$ should only be affected by vacuum fermion loops
(dynamical configurations) at the percentage level, Ref.~\cite{Burgio:2012ph}. Hence, when using 
costly dynamical overlap configurations we do not expect much difference for the dynamical mass function.

\begin{figure}[t]
 \centering
 \includegraphics[angle=0,width=.98\linewidth]{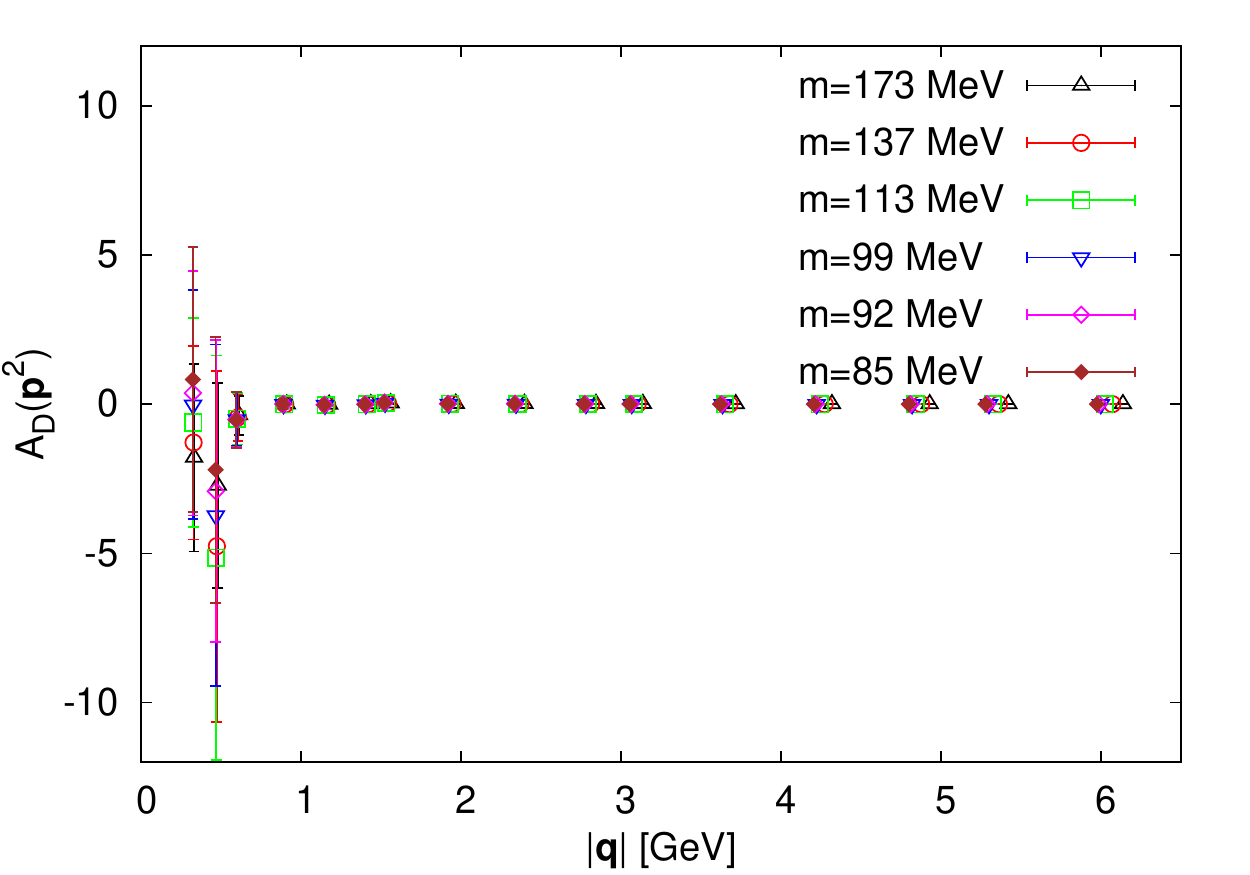}
 \caption[Temporal component for full case II]{\sl
  Mixed dressing function $A_{\textsc{d}}(\bp)$ for several current quark masses.}
 \label{mixed-component}
 \end{figure}

The mixed dressing function $A_{\textsc{d}}(\bp)$ seems to vanish for all current quark masses, see Fig.~\ref{mixed-component}. For small
momenta the error bars are too large to make a precise statement. 
 
\begin{figure}[t]
 \centering
 \includegraphics[angle=0,width=.98\linewidth]{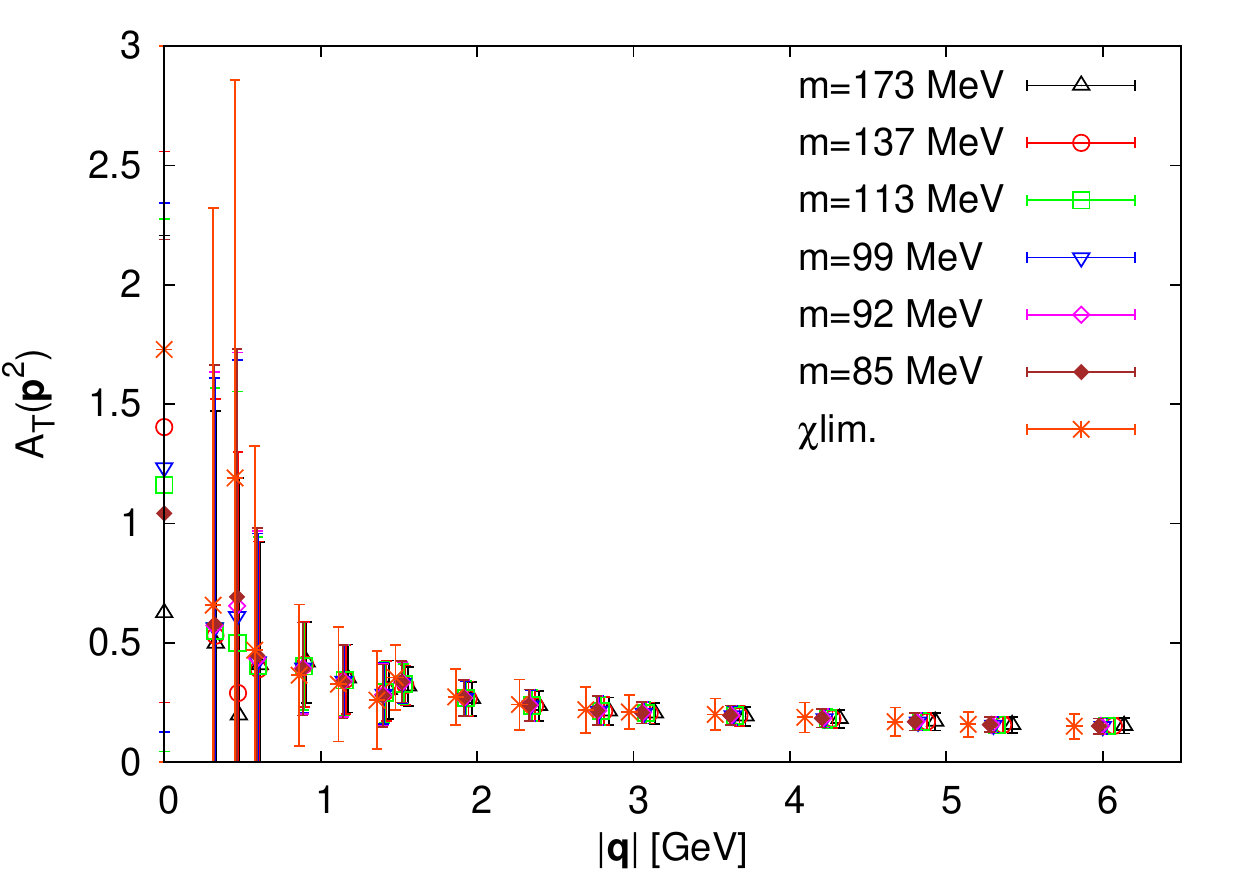}
 \caption[Temporal component for full case II]{\sl
  Temporal dressing function $A_{\textsc{t}}(\bp)$ for several current quark masses and in the chiral limit.}
 \label{temporal-component}
 \end{figure}

Finally, we comment on the temporal dressing function 
$A_{\textsc{t}}(\bp)$, which is only well-defined when the residual gauge invariance with respect to space independent gauge transformations is fixed.
However, it is not influenced by the Coulomb gauge condition. 
The additional gauge freedom is fixed via the Integrated Polyakov gauge. 
In this gauge $A_{\textsc{t}}(\bp)$ goes to small non-zero values for large momenta. For small momenta the error bars are too large to make a 
precise statement, see Fig.~\ref{temporal-component}. However, a divergent behavior, as for the
dressing functions fixed by Coulomb gauge, does not seem to take place. 
We note that the qualitative behavior is consistent with the findings in Ref.~\cite{Burgio:2012ph}. 
For large and intermediate momenta no mass dependence is seen. For small momenta
there is a tendency that smaller masses approach larger values. However, one has to be 
careful, since the error bars are large in this region to make a precise statement. The 
temporal dressing function should also be clarified in the continuum approach, which,
until now, has not been considered. 

 We stress that for the confinement properties the dressing function $A_{\textsc{s}}(\bp)$ is the quantity of central interest. 
This issue will be clarified in the next section.

\subsection{Comparison with propagator on stout-smeared configurations}
\label{stout-compare}
In order to understand the role of smearing for the quark propagator in Coulomb gauge,
we compare the dressing functions on stout-smeared configurations with the unsmeared case. 
We use the same mass parameters $m_0$ in both cases. We do not interpolate
between the mass values, as done in Ref.~\cite{Zhang:2009jf}, where the effect of smearing
on the Landau gauge quark propagator was under consideration. 

\begin{figure}[t]
 \centering  
 \includegraphics[angle=0,width=.98\linewidth]{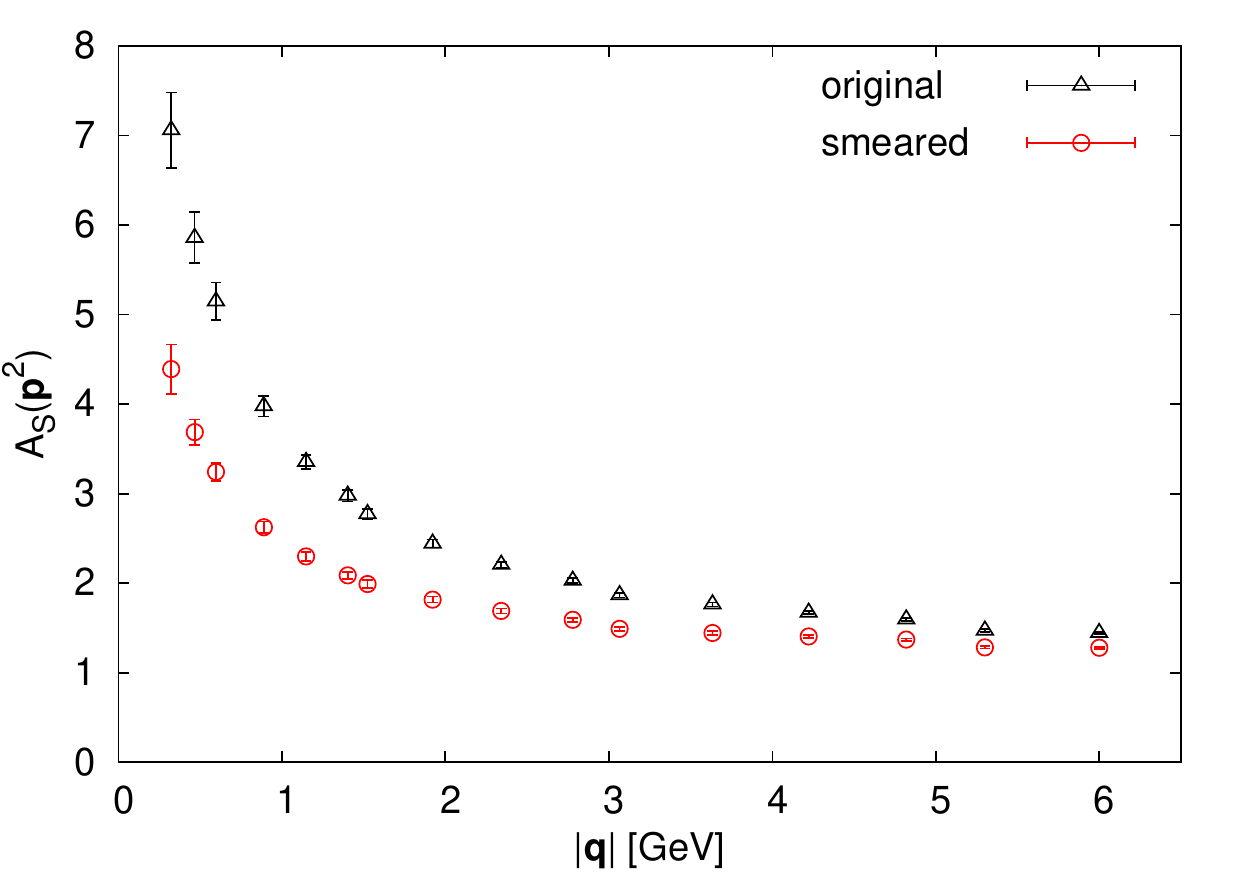}
 \caption[Spatial component for full case]{\sl
 Comparison of spatial component $A_{\textsc{s}}(\bp)$ with three levels of stout smeared and unsmeared configurations for 
 the mass value $m=99$ MeV.}
 \label{spatial-component-smeared}
 \end{figure}

\begin{figure}[t]
 \centering  
 \includegraphics[angle=0,width=.98\linewidth]{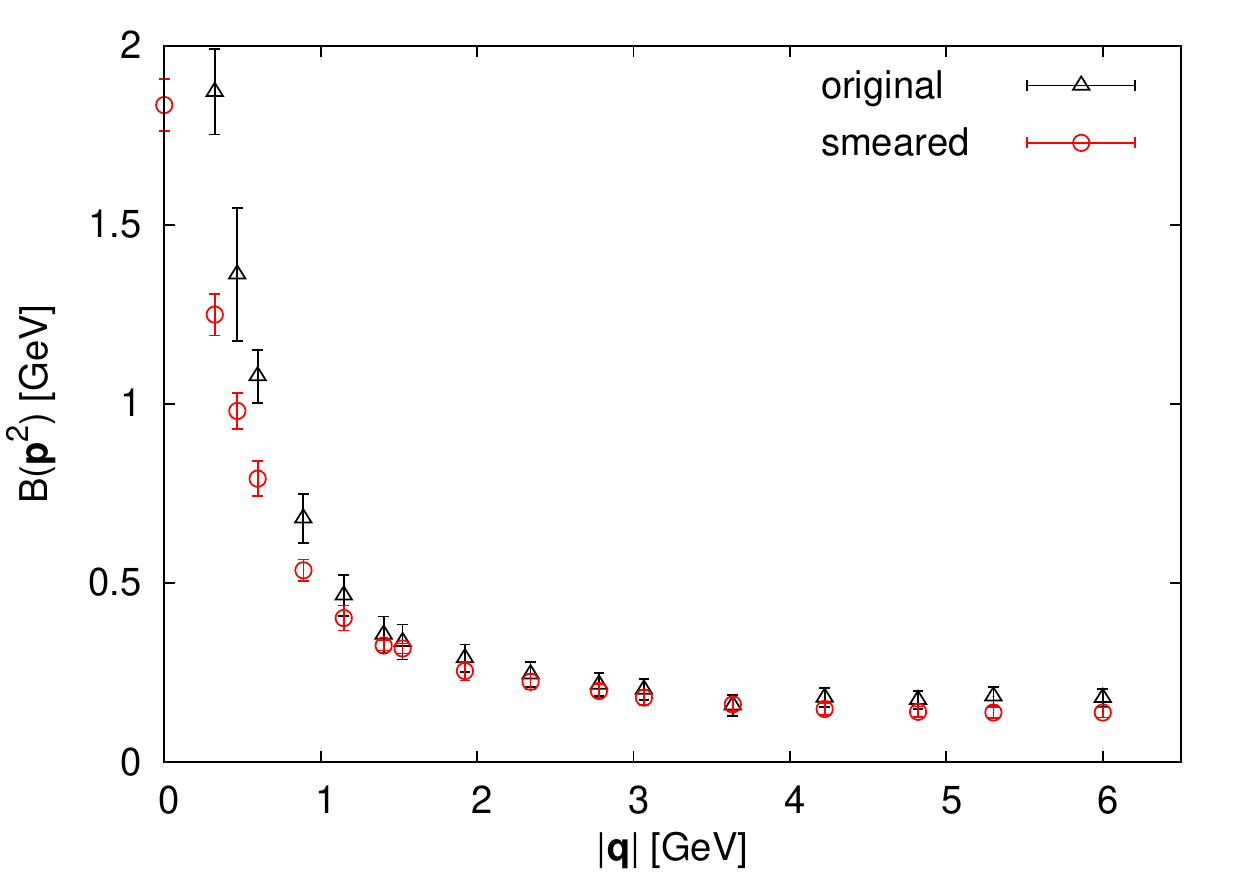}
 \caption[Spatial component for full case]{\sl
 Comparison of scalar component $B(\bp)$ with three levels of stout smeared and unsmeared configurations for 
 the mass value $m=99$ MeV.}
 \label{scalar-component-smeared}
 \end{figure}

\begin{figure}[t]
 \centering
 \includegraphics[angle=0,width=.98\linewidth]{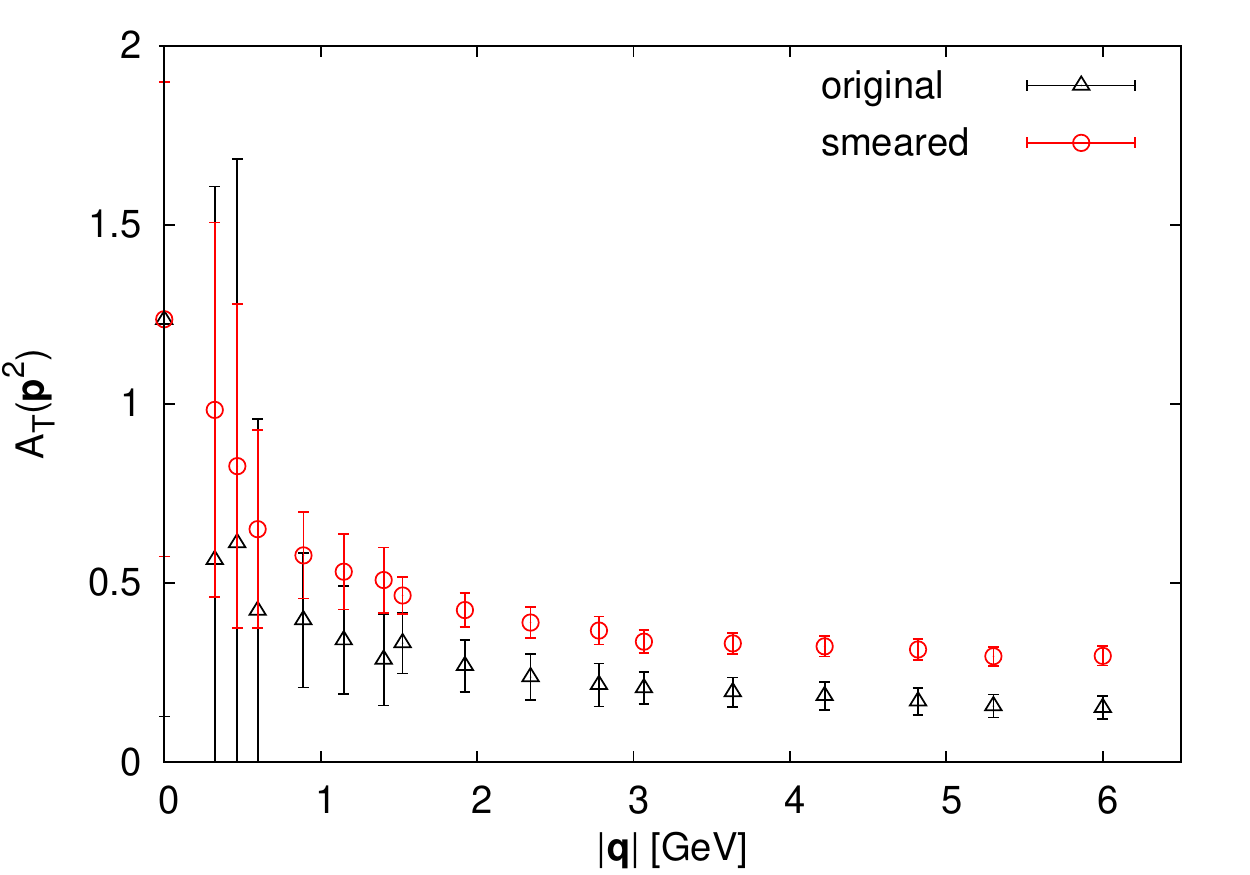}
 \caption[Temporal component for full case II]{\sl
  Temporal dressing function $A_{\textsc{t}}(\bp)$ with three levels of stout smeared and unsmeared configurations for 
 the mass value $m=99$ MeV.}
 \label{temporal-component-smeared}
 \end{figure}

All dressing functions show the same qualitative behavior as in the unsmeared case. The functions
$A_{\textsc{s}}(\bp)$, $B(\bp)$ and $A_{\textsc{t}}(\bp)$ still increase for small momenta, see 
Figs.~\ref{spatial-component-smeared}, \ref{scalar-component-smeared}, \ref{temporal-component-smeared}. 
However, the dressing functions approach different low-momentum values as in the unsmeared case. 
The functions $A_{\textsc{s}}(\bp)$ and $B(\bp)$ reach \textit{smaller} values 
for low-momenta, whereas $A_{\textsc{t}}(\bp)$ reaches \textit{larger}
values. In the mid-momentum regime   
$A_{\textsc{s}}(\bp)$ gives smaller values as compared to the unsmeared case, whereas the function $B(\bp)$ gives nearly
the same values as in the unsmeared case. The scalar dressing functions start to deviate from each other around $1$ GeV. 
For $A_{\textsc{t}}(\bp)$ the values are slightly larger in the whole momentum regime. 

We also observe, that the error bars
for the IR values for all dressing functions are drastically reduced compared to the unsmeared case. This is especially
important for the temporal dressing function $A_{\textsc{t}}(\bp)$, see Fig.~\ref{temporal-component-smeared}, which
gives now a clearer IR behavior, as for the unsmeared case, see Fig.~\ref{temporal-component} for comparison.
From this result we can conclude that the mass dependence of the temporal dressing function $A_{\textsc{t}}(\bp)$ is small. The fact, 
that the error bars are reduced for the IR values, seems to be a direct consequence of the smoothing procedure. 

\begin{figure}[t]
 \centering  
 \includegraphics[angle=0,width=.98\linewidth]{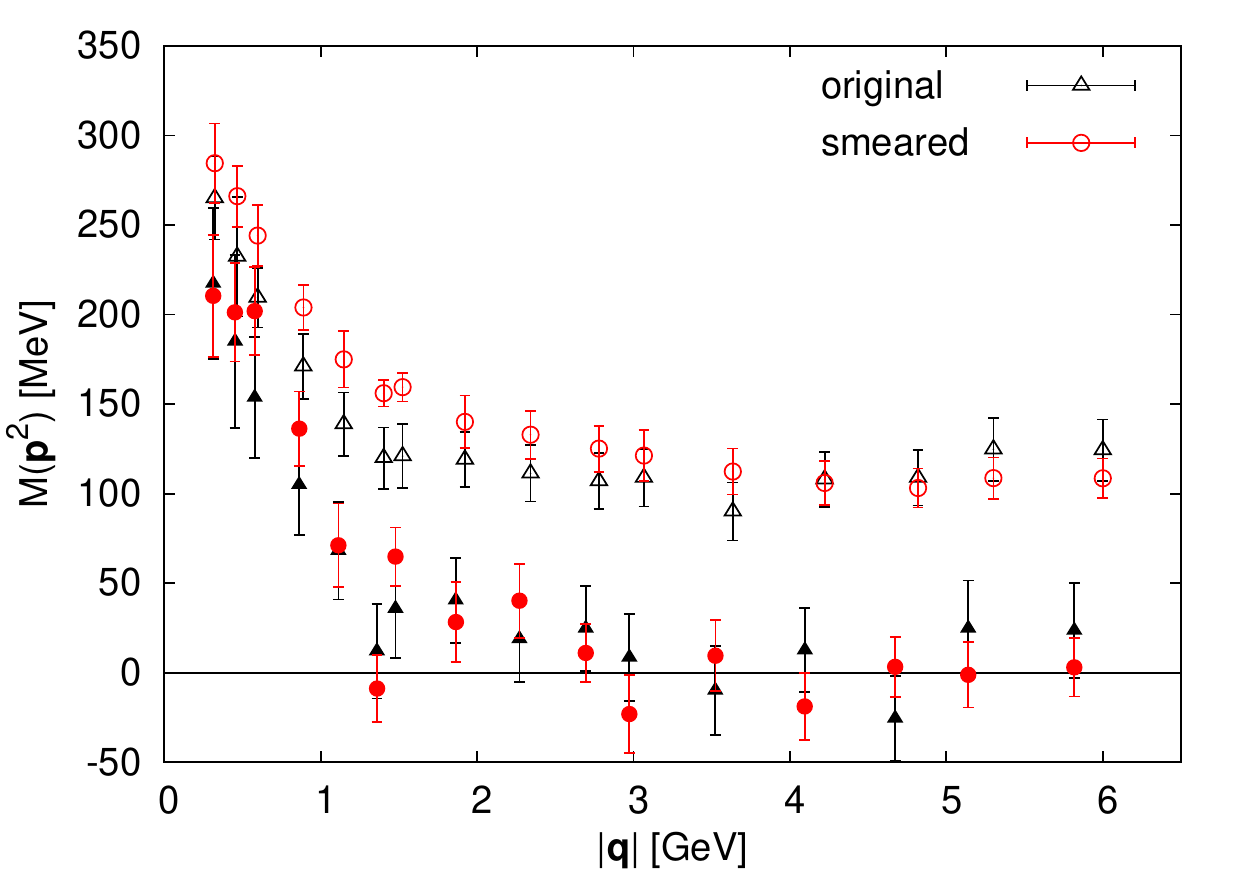}
 \caption[Spatial component for full case]{\sl
 Dynamical mass function $M(\bp)$ with three levels of stout smeared and unsmeared configurations for 
 the mass value $m=99$ MeV (open symbols) and in the chiral limit (full symbols).}
 \label{dynamical-mass-smeared}
 \end{figure}

In Fig.~\ref{dynamical-mass-smeared} we compare $M(\bp)$ for smeared and unsmeared configurations for $m=99$ MeV 
and in the chiral limit. 
In contrast to the dressing functions, the dynamical quark mass $M(\bp)$ agrees quantitatively, within the error bars, for smeared and unsmeared configurations. 
In the smeared case it reaches slightly larger infrared values for finite current quark mass $m$,
which is due to the fact, that $A_{\textsc{s}}(\bp)$ in the IR is smaller than for unsmeared configurations.
Clearly, as pointed out in Ref.~\cite{Zhang:2009jf}, the overlap mass parameter $m_0$ on the smeared configurations gives a
different renormalized quark mass. However, in the chiral limit smeared and unsmeared propagators are nearly indistinguishable, and a constituent mass 
of around $300$ MeV is reached. This is a remarkable result: although scalar and vector dressing functions give quite different 
values for the smeared propagator, the dynamical quark mass $M(\bp)$ is not really affected by
the smearing procedure. The differences cancel each other in the ratio Eq.~(\ref{dynamical-mass}).

\section{Confinement in Coulomb Gauge and Artificial Chiral Symmetry Restoration}
\label{Chap-Low-Mode}
We clarify how confinement is related to the infrared divergent quark dressing functions in Coulomb gauge. Such a connection has first been
discovered in a continuum Coulomb gauge model, Refs.~\cite{Adler:1984ri, Alkofer:1988tc, Wagenbrunn:2007ie}. 
Only recently it has been explored in lattice calculations as well, see Refs.~\cite{Burgio:2012ph, Burgio:2013mx}.

Integrating the free (Euclidean) quark propagator over the component $p_4$ gives the free static quark propagator
\begin{align}
 S(\bp) = \int \frac{dp_4}{2\pi} \frac{1}{i \boldsymbol{\gamma} \cdot \bp + i \gamma_4 p_4 + m_0} = \frac{m_0 - i \boldsymbol{\gamma} \cdot \bp}{2 \omega(\bp)} \; ,
\end{align}
where in the denominator the free-particle dispersion relation $\omega(\bp) = \sqrt{\bp^2+m^2 }$ appears. 
Performing the $p_4$-integration for the interacting quark propagator $S(\bp,p_4)$, Eq.~(\ref{prop-parameterization}), yields 
\begin{align}
\label{static-prop}
 S(\bp) = \frac{B(\bp) - i \boldsymbol{\gamma} \cdot \bp \, A_{\textsc{S}}(\bp)}{2 \omega(\bp)} ,
\end{align}
with the dispersion relation of a confined quark
\begin{align}
\label{disp1}
 \omega(\bp) = A_{\textsc{T}}(\bp) \sqrt{\bp^2 A^2_{\textsc{S}}(\bp)+B^2(\bp)} \; .
\end{align}
If scalar and vector dressing functions, $B(\bp)$ and $A_{\textsc{S}}(\bp)$, respectively, are IR divergent, 
the energy dispersion $\omega(\bp)$ is IR divergent as well. 
This gives rise to the physical picture of quark confinement: the excitation energy of a single quark is infinite. 
It has to be stressed that such a mechanism of quark confinement via an IR divergent quark dispersion relation is special for Coulomb gauge and absent, for instance, in Landau gauge.

Using the dynamical quark mass $M(\bp)$, Eq.~(\ref{dynamical-mass}), we can rewrite Eq.~(\ref{disp1}) as
\begin{align}
\label{disp2}
 \omega(\bp) = A_{\textsc{T}}(\bp) A_{\textsc{S}}(\bp)\sqrt{\bp^2 + M^2(\bp)} \; .
\end{align}
Supposing that $M(\bp)$ and $A_{\textsc{T}}(\bp)$ are constant for $|\bp|\rightarrow 0$, the IR-divergence of $\omega(\bp)$ is 
fully included in the infrared divergence of $A_{\textsc{S}}(\bp)$. 
Hence, the vector dressing function $A_{\textsc{S}}(\bp)$ is crucial to link the confinement properties of quarks 
to chiral symmetry breaking. In Ref.~\cite{Burgio:2012ph} first indications of an IR-diverging quark 
dispersion relation $\omega(\bp)$ on the lattice have been observed.

Now an interesting question arises: if removing the chiral condensate from the quark propagator,
is the energy dispersion $\omega(\bp)$, Eq.~(\ref{disp1}), still infrared divergent?
If yes, then confinement is intact, although chiral symmetry is artificially restored in the vacuum.
We will analyze this question numerically. Via the Banks-Casher relation~\cite{Banks:1979yr} the 
chiral condensate is connected to the low-lying eigenvalues of the Dirac operator. These low-lying eigenvalues  
are removed via the
prescription
\begin{align}\label{eq:truncation}
 S^{k}_{\textsc{trunc}} = S_{\textsc{full}} - \sum_{i=1}^{k} \frac{1}{\lambda_i} |v_i\rangle \langle v_i | \; ,
\end{align}
with $S^k_{\textsc{trunc}}$ and $S_{\textsc{full}}$ denoting low-mode truncated and full propagators, respectively. Here $\lambda_i$ are
the eigenvalues, $|v_i\rangle$ the eigenvectors of the overlap Dirac operator and $k$
denotes the number of truncated modes. For instance, $k=16$ means subtracting the 16 lowest modes from the quark propagator. 
For our $20^4$ lattice we use truncation steps from 16 to 256, corresponding to removing all eigenvalues of the massless Dirac operator spectrum up to values of $6-138$ MeV, see Tab.~\ref{tab:trunclevels} for the single steps. After removing enough modes, the 
quark condensate is zero. The truncated propagators $S^k_{\textsc{trunc}}$
are evaluated for all six mass values and the chiral limit is considered as well. 
\begin{table}[htdp]
\begin{center}
\begin{tabular}{c|c}
$k$ & cutoff [MeV] \\\hline
16  & $6\pm	    3$ \\
32  & $14\pm	3 $\\
64  & $29\pm	3$ \\
96  & $46\pm	4$ \\
128 & $63\pm	5$ \\
256 & $138\pm	6$ 
\end{tabular}
\end{center}
\caption{The truncation steps $k$ and the corresponding eigenvalue cutoffs of the massless Dirac operator at which we evaluate the quark propagator. The cutoff values are averages over all configurations and the errors are standard deviations.}
\label{tab:trunclevels}
\end{table}%

It is important to note that while the truncation (\ref{eq:truncation}) leaves unitarity intact, by construction it violates the locality of the Dirac operator to some extent. In \cite{Schrock:2013rea} it has been shown, however, that the violation of locality of a truncated Wilson type Dirac operator is many orders of magnitude smaller than the ultra local contribution.  Moreover, in Ref.~\cite{Hernandez:1998et} it has been demonstrated that, at finite lattice spacing, the (untruncated) overlap operator reveals nonlocal contributions which fall of exponentially as a function of the distance.
Therefore we are confident that the violation of locality resulting from the low mode truncation Eq.~(\ref{eq:truncation}) does not affect our results.

After chiral symmetry restoration, the quark condensate 
$\langle \overline{\psi} \psi \rangle$ and therefore the dynamical quark mass $M(\bp)$ and scalar dressing function $B(\bp)$ vanish in the chiral limit.
The quark dispersion relation $\omega(\bp)$, Eq.~(\ref{disp2}), then reads
\begin{align}
\label{disp-symm-restore}
 \omega_{\textsc{trunc}}(\bp) = A_{\textsc{T}}(\bp) A_{\textsc{S}}(\bp) \, |\bp| \; . 
\end{align}
The important question now is, what happens with the vector dressing function $A_{\textsc{S}}(\bp)$ after chiral symmetry restoration. 
This goes in hand with the question, whether confinement survives artificial chiral symmetry restoration\footnote{It can be speculated that $A_{\textsc{S}}(\bp)$ serves an an order parameter of confinement in 
Coulomb gauge. This issue will be explored further in future studies, for instance, by studying the 
quark propagator in the deconfinement region.}.

We note that in Ref.~\cite{Glozman:2007tv}, within a continuum Coulomb gauge model, it is demonstrated that with removing all the momenta relevant
for the chiral condensate, confinement is still intact. It is shown that the vector dressing function $A_{\textsc{S}}(\bp)$ is still infrared divergent, while the scalar
dressing function $B(\bp)$ goes to zero. In the next section we will explore numerically, if it is possible to generalize this picture to full QCD.

In the remainder of this section we clarify how the dynamical quark mass $M(\bp)$, a gauge variant quantity,
is connected to the chiral condensate $\langle \overline{\psi}(\bx) \psi(\bx) \rangle$ which, in contrast, is  gauge invariant. 
 
Using Eq.~(\ref{static-prop}) the chiral condensate, which is given as the trace of the static quark propagator, can be expressed as
 \begin{align}
 \label{chiral-cond}
  \langle \overline{\psi}^a(\bx) \psi^a(\bx) \rangle = - 2 \, N_{\textsc{C}} \int \frac{d^3 p}{(2\pi)^3} \, \frac{M(\bp)}{A_{\textsc{t}}(\bp) \sqrt{\bp^2+M^2(\bp)}} \; .  
 \end{align}
We note that the integral on the right-hand side is only well-defined in the chiral limit. From this 
formula it follows that the chiral condensate vanishes when $M(\bp) =0$, assuming that $A_{\textsc{t}}(\bp)$ stays finite. 

The expression (\ref{chiral-cond}) could be used to relate Dirac eigenvalues, used on the
lattice to artifically remove the chiral condensate, and quark momenta, used in functional methods like Dyson-Schwinger equations for this purpose.
For instance, it is still not fully understood why, after only a small amount of low-modes are removed, where the chiral condensate is not yet zero, 
mesons already appear in chiral multiplets. New insights on this issue could be gained by comparing these two approaches\footnote{The procedure would be as follows: after removing a small amount of low-lying modes, the constituent quark mass $M(0)$ has a certain non-zero value below the QCD value around $300$ MeV. 
In a functional approach the integration momenta $|\bp| < |\bp|_{\textsc{crit}}$ in Eq~.(\ref{chiral-cond}) should be excluded and $|\bp|_{\textsc{crit}}$ adjusted to get the
same value for $M(0)$. Then the meson spectrum should be evaluated by solving the Bethe-Salpeter equation.}. 

\subsection{Numerical results}
\label{Chap-Low-Mode-Results}
For each current quark mass $m$ we evaluate the dressing functions for all six truncation steps. 
The dressing functions are shown in the chiral limit. The mass function $M(\bp)$ and 
scalar dressing function $B(\bp)$ are also shown for the current quark mass $m=99$ MeV.  
Our conclusions apply for all mass values.  

\begin{figure}[t]
 \centering
 \includegraphics[angle=0,width=.98\linewidth]{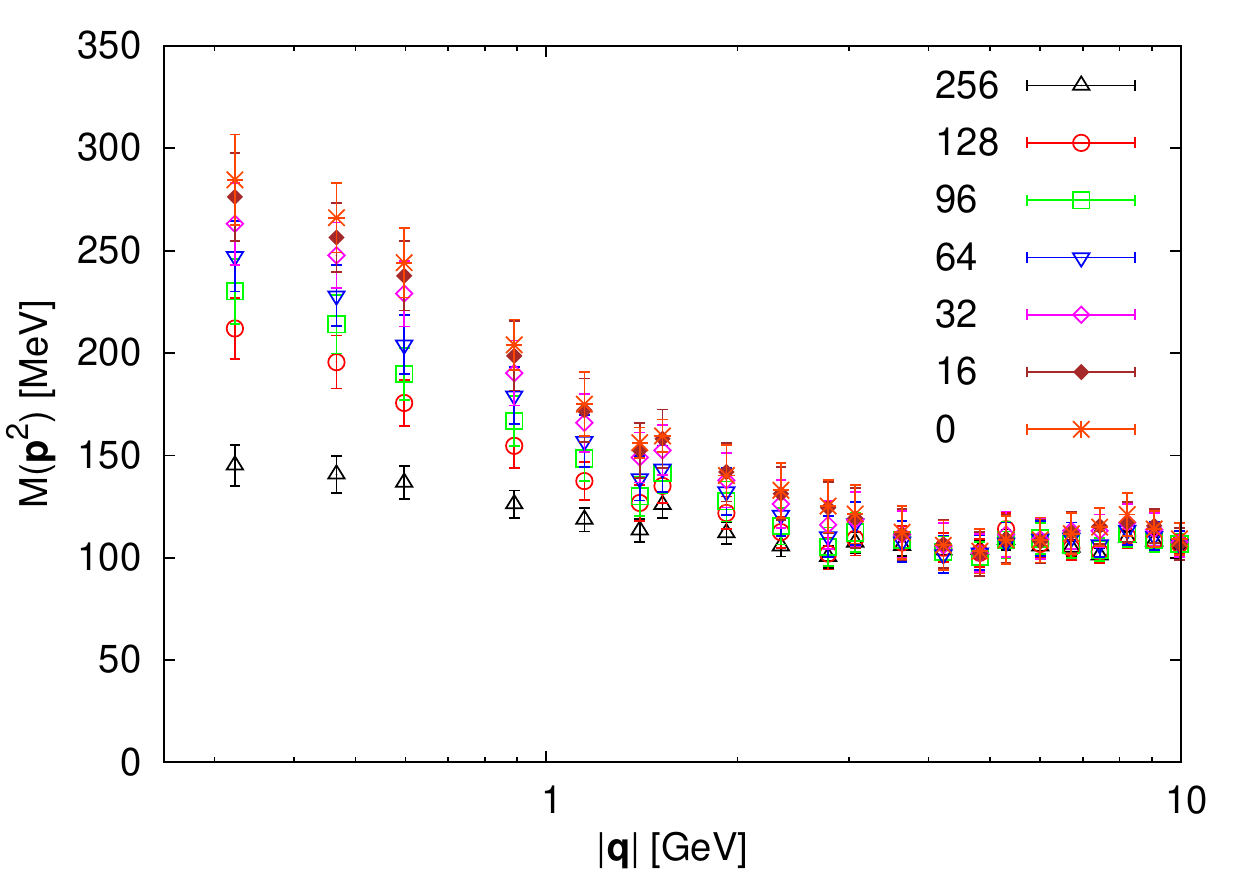}
 \caption[Mass function truncation chiral limit]{\sl
Dynamical mass function $M(\bp)$ for the untruncated case and for six truncation steps for the current quark mass $m=99$ MeV.}
 \label{M-trunc-mass}
 \end{figure}

\begin{figure}[t]
 \centering
 \includegraphics[angle=0,width=.98\linewidth]{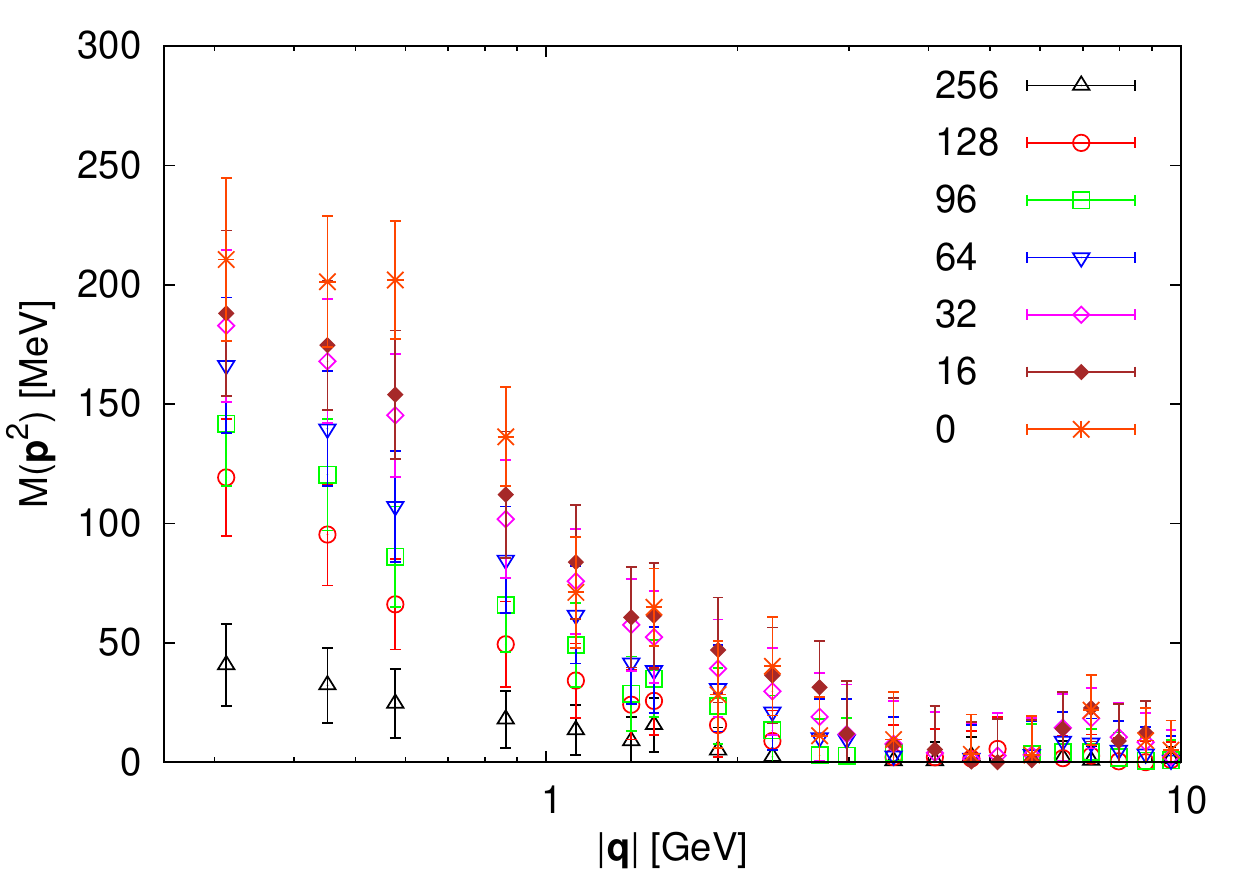}
 \caption[Mass function truncation chiral limit]{\sl
Dynamical mass function $M(\bp)$ for the untruncated case and for six truncation steps in the chiral limit.}
 \label{M-trunc-chiral}
 \end{figure}

We start with discussing the dynamical quark mass $M(\bp)$, Fig.~\ref{M-trunc-mass}. 
As expected from chiral symmetry considerations, it decreases for increasing number of truncated Dirac eigenmodes. 
Already after the first truncation step ($k=16$), $M(\bp)$ has lost some IR strength. 
After removing $k=256$ modes the function appears close to the current quark mass, $M(\bp) \approx m$. 
In the chiral limit this results in $M(\bp)$ approaching a small (but non-zero) IR value, see Fig.~\ref{M-trunc-chiral}, after $k=256$ low modes have been removed. 
Correspondingly, the chiral condensate is small but not yet zero at that stage.

\begin{figure}[t]
 \centering
 \includegraphics[angle=0,width=.98\linewidth]{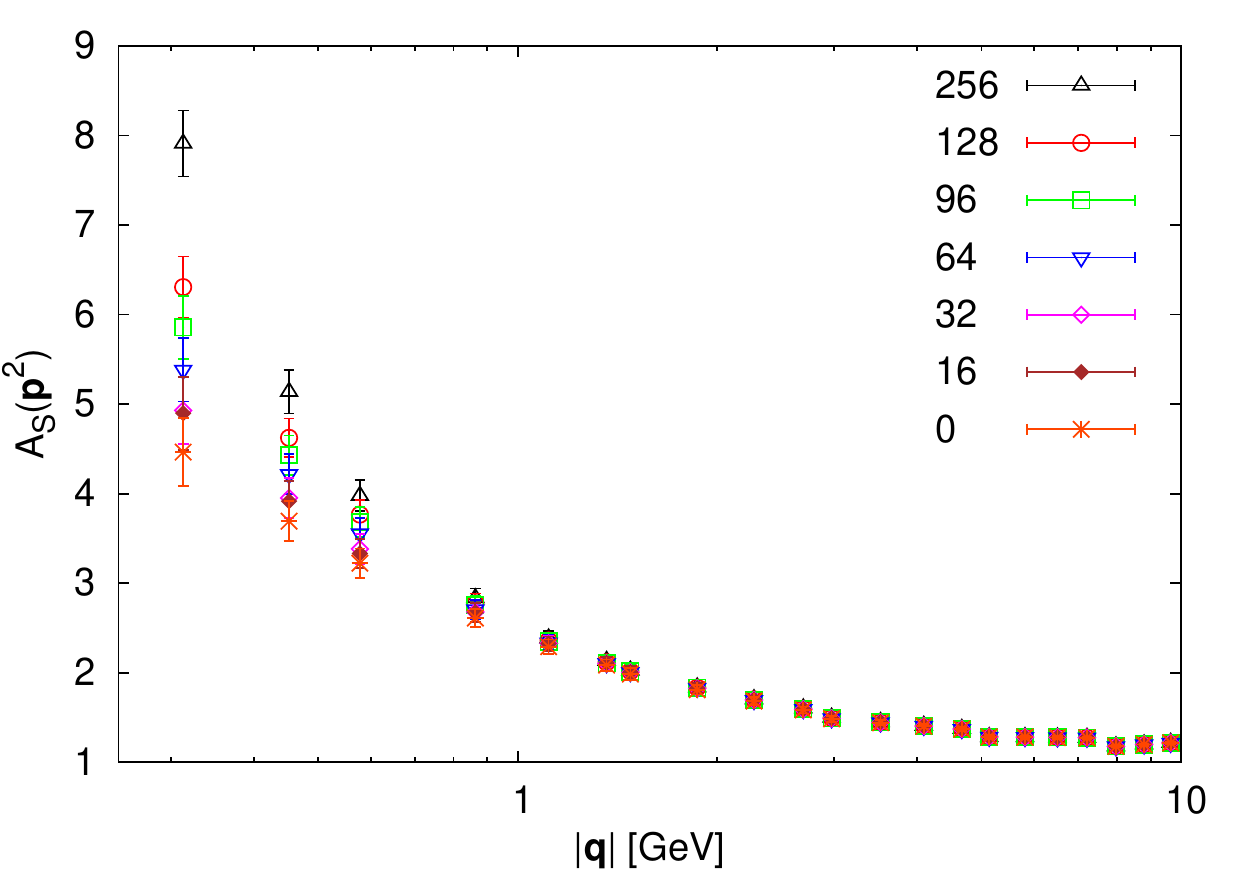}
 \caption[Spatial dresssing truncation]{\sl
Spatial dressing function $A_{\textsc{S}}(\bp)$ for the untruncated case and for six truncation steps in the chiral limit.}
 \label{A-trunc}
 \end{figure}

The spatial dressing function $A_{\textsc{S}}(\bp)$, however, does not decrease after removing the low-lying modes, see Fig.~\ref{A-trunc}.
Instead, the small momentum values even \emph{increase} after removing the low-lying eigenvalues.
While for large and intermediate momenta (around 1 GeV) the values for each truncation
step lie on top of each other. 
The former is one of the crucial observations of this study. It is a clear indication that an infrared divergent dispersion relation $\omega(\bp)$, Eq.~(\ref{disp-symm-restore}), 
survives the artificial chiral symmetry restoration.\footnote{In a next step the IR behavior of the spatial dressing function should be analyzed in detail, 
for instance,
it should be checked, if it obeys a power-law $A_{\textsc{S}}(\bp \rightarrow 0) \sim |\bp|^{-\alpha}$, and understood, why
after truncation, the value for $\alpha$ becomes even larger. 
For such an analysis the continuum Coulomb gauge model
could be well-suited.}

\begin{figure}[t]
 \centering
 \includegraphics[angle=0,width=.98\linewidth]{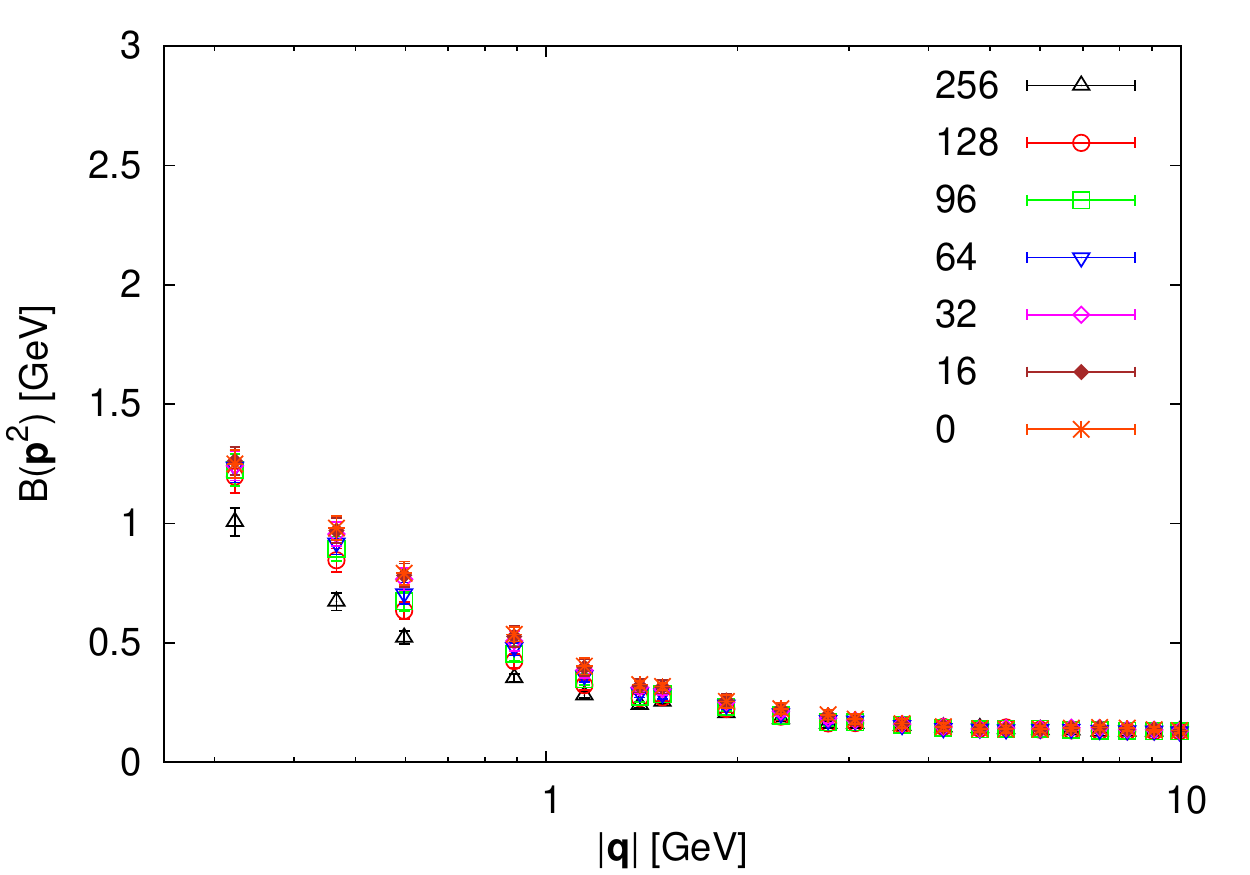}
 \caption[Spatial dresssing truncation]{\sl
Scalar dressing function $B(\bp)$ for the untruncated case and for six truncation steps for the current quark mass $m=99$ MeV.}
 \label{B-trunc-mass}
 \end{figure}

\begin{figure}[t]
 \centering
 \includegraphics[angle=0,width=.98\linewidth]{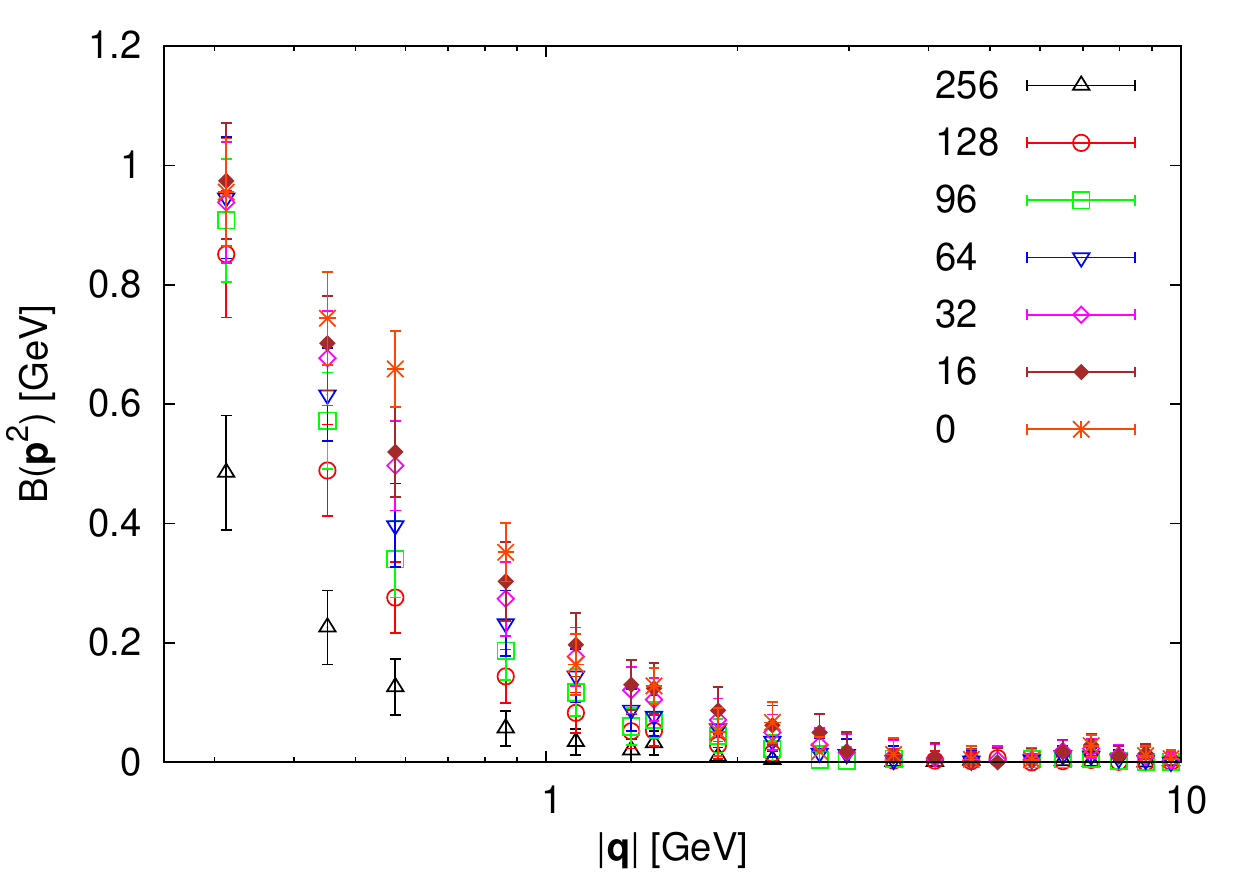}
 \caption[Spatial dresssing truncation]{\sl
Scalar dressing function $B(\bp)$ for the untruncated case and for six truncation steps in the chiral limit.}
 \label{B-trunc-chiral}
 \end{figure}

The scalar dressing function $B(\bp)$ decreases under low-mode truncation, see Fig.~\ref{B-trunc-mass}, consistent
with chiral symmetry considerations.
However, for non-vanishing quark mass $B(\bp)$ increases also at the largest truncation level $k=256$ quite strongly towards the IR.
This behavior can be understood by considering the mean-field Dyson-Schwinger equations of the truncated theory (Ref.~\cite{Glozman:2007tv}): 
the bare quarks still interact with gluons via the color-Coulomb potential. 
Only in the chiral limit $B(\bp)$ vanishes after removing enough low modes, since the Dirac structure $\mathds{1}$ is then not allowed.
In our case, after removing $k=256$ modes, $B(\bp)$ has lost more than the half of its IR strength, see Fig.~\ref{B-trunc-chiral}. 
But it is still non-zero, which again confirms that the chiral condensate is non-zero also at the largest 
truncation level.\footnote{In Coulomb gauge the infrared divergencies, which occur for spatial and scalar dressing functions,
cancel each other in the dynamical mass function and the observable quantites. Since
in the chirally symmetric phase the dynamical quark mass $M(\bp)$ wants to approach the current quark mass $m$ 
in the IR, 
$B(\bp)$ and $A_{\textsc{s}}(\bp)$ have to change their IR behavior, so that $M(\bp)$ is still finite.}

\begin{figure}[t]
 \centering
 \includegraphics[angle=0,width=.98\linewidth]{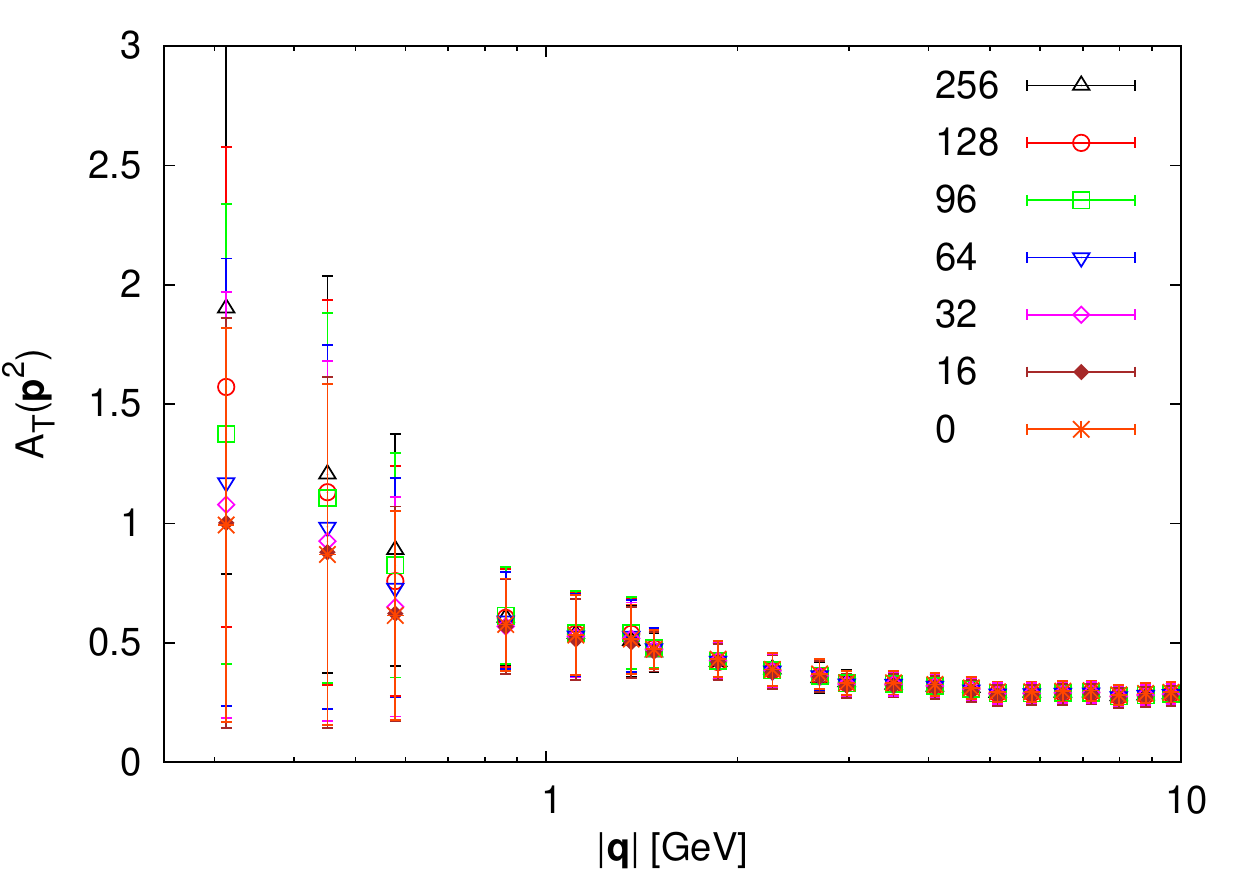}
 \caption[Spatial dresssing truncation]{\sl
Temporal dressing function $A_{\textsc{T}}(\bp)$ for the untruncated case and for six truncation steps in the chiral limit.}
 \label{At-trunc}
 \end{figure}

The temporal dressing function $A_{\textsc{T}}(\bp)$ is also allowed in the chirally symmetric scenario. 
It stays non-zero after truncation as shown in Fig.~\ref{At-trunc}. It behaves similar to the spatial dressing function $A_{\textsc{S}}(\bp)$. It does not change under truncation
for low and intermediate momenta. For small momenta it seems to increase, although the gauge noise is too large
to make a conclusive statement. 

The mixed component $A_{\textsc{D}}(\bp)$, which does not occur in the untruncated case, is still
vanishing for the truncated case and therefore not shown here. 

Let us underline the main outcome of this numerical study. After removing the low-lying eigenmodes from the quark propator,
which are responsible for the formation of the chiral condensate,
the Lorentz-vector dressing function, which enters the quark dispersion relation $\omega(\bp)$, Eq.~(\ref{disp-symm-restore}),
stays intact and even becomes stronger divergent in the IR.   
This gives the Coulomb gauge explanation of confinement after artifical chiral symmetry restoration: 
due to the persistence of the vector dressing function $A_{\textsc{S}}(\bp)$, the quark dispersion relation 
$\omega(\bp)$, Eq.~(\ref{disp-symm-restore}), is still IR divergent and therefore 
a single quark is still absent from the spectrum. This conclusion supports recent hadron spectroscopy studies,
see Refs.~\cite{Lang:2011vw, Glozman:2012fj, Denissenya:2014poa}, where the symmetries of hadrons are analyzed 
under low-mode truncation. The fact that confinement is still intact can be judged from the exponential
decay signal of hadron correlators. 
Here we have found a more physical, though gauge dependent, explanation for this phenomenon. 

 
\section{Summary and Conclusions}
\label{Chap-Summary}
A first detailed analysis of the overlap quark propagator in Coulomb gauge has been presented. 
The use of chirally symmetric lattice fermions makes the results highly predictive,
since no improvement techniques have to be implemented, which could spoil the result. The drawback is the immense 
computational cost. To reach the small momentum region a large lattice has to be chosen,
which is in our case a $20^4$ lattice with $a=0.2$~fm. 

We observe a clear indication of dynamical mass generation in the
quark propagator dressing functions. Scalar and vector dressing functions and therefore the dynamical quark mass 
start to increase in the momentum region around $1-2$ GeV. The constituent quark mass in the chiral limit is around $300$ MeV, 
which is expected to change only slightly for dynamical configurations. Moreover, we see hints that scalar and vector dressing functions, and therefore
the energy dispersion, are diverging in the infrared limit. However, such a statement can only be confirmed by taking 
the continuum limit. Moreover, for a suitable parameterization of the dynamical mass function, put forward in Ref.~\cite{Burgio:2012ph},
the continuum limit is of great importance. Such a parameterization would be very helpful for continuum
Coulomb gauge studies. 

We also studied the quark propagator after applying three levels of stout smearing to the gauge field configurations. 
Although the dressing functions show the same qualitative behavior as in the unsmeared case, 
they reach different small momentum values. The interesting observation is that 
the dynamical mass function is not affected by the smearing procedure and is only shifted to slightly larger values.

In the second part of the paper we explored the interrelation between confinement and chiral symmetry
breaking. Coulomb gauge is very appealing for this purpose, since an energy dispersion can be formulated,
which relates the confinement properties of quarks to the divergence of the dressing functions. 
We created a scenario where the low modes of the Dirac operator are removed from the valence quark propagators and thereby the 
chiral condensate is suppressed. We found that, in contrast to the scalar dressing function $B(\bp)$, 
the vector dressing function $A_{\textsc{s}}(\bp)$ stays intact in the chiral limit and even seems to diverge 
stronger (again, for a conclusive result the continuum behavior has to be explored). Therefore the
quark dispersion relation $\omega(\bp)$ can still be infrared divergent. 
It is the vector dressing function which 
is responsible that a single quark is still confined in the hadron. We stress that to our knowledge
such a direct connection between confinement and chiral symmetry breaking is only possible in Coulomb gauge.

Moreover, the relation between the gauge dependent dynamical mass function and the gauge invariant chiral condensate has been given: the vanishing of the former implies the vanishing of the latter. 

The Coulomb gauge dispersion relation for confined 
quarks offers the possibility to study several interesting questions regarding the relation between confinement and 
chiral symmetry breaking, like, e.g., how do the dressing functions behave if instantons are removed
from the quark propagator, or in the case of adjoint quarks? What happens at finite temperature?
We expect that Coulomb gauge could shed light on these questions and help to understand the interrelation of confinement and chiral symmetry breaking.

\begin{acknowledgements}
We thank Y.~Delgado for being a co-worker at the early stages of the project. 
We thank C.~B.~Lang for sharing his overlap fermion codes with us.
Discussions with G.~Burgio and L.~Glozman are greatly acknowledged.  
M.~P. acknowledges support by the Austrian Science Fund (FWF)
through the grant P26627-N27. The calculations have been performed on clusters 
at ZID at the University of Graz and 
at the Graz University of Technology. 
\end{acknowledgements}

\bibliographystyle{apsrev4-1}
\bibliography{Overlap-Bib}

\begin{thebibliography}{46}%
\makeatletter
\providecommand \@ifxundefined [1]{%
 \@ifx{#1\undefined}
}%
\providecommand \@ifnum [1]{%
 \ifnum #1\expandafter \@firstoftwo
 \else \expandafter \@secondoftwo
 \fi
}%
\providecommand \@ifx [1]{%
 \ifx #1\expandafter \@firstoftwo
 \else \expandafter \@secondoftwo
 \fi
}%
\providecommand \natexlab [1]{#1}%
\providecommand \enquote  [1]{``#1''}%
\providecommand \bibnamefont  [1]{#1}%
\providecommand \bibfnamefont [1]{#1}%
\providecommand \citenamefont [1]{#1}%
\providecommand \href@noop [0]{\@secondoftwo}%
\providecommand \href [0]{\begingroup \@sanitize@url \@href}%
\providecommand \@href[1]{\@@startlink{#1}\@@href}%
\providecommand \@@href[1]{\endgroup#1\@@endlink}%
\providecommand \@sanitize@url [0]{\catcode `\\12\catcode `\$12\catcode
  `\&12\catcode `\#12\catcode `\^12\catcode `\_12\catcode `\%12\relax}%
\providecommand \@@startlink[1]{}%
\providecommand \@@endlink[0]{}%
\providecommand \url  [0]{\begingroup\@sanitize@url \@url }%
\providecommand \@url [1]{\endgroup\@href {#1}{\urlprefix }}%
\providecommand \urlprefix  [0]{URL }%
\providecommand \Eprint [0]{\href }%
\@ifxundefined \urlstyle {%
  \providecommand \doi  [0]{\begingroup \@sanitize@url \@doi}%
  \providecommand \@doi [1]{\endgroup \@@startlink {\doibase
  #1}doi:\discretionary {}{}{}#1\@@endlink }%
}{%
  \providecommand \doi  [0]{doi:\discretionary{}{}{}\begingroup
  \urlstyle{rm}\Url }%
}%
\providecommand \doibase [0]{http://dx.doi.org/}%
\providecommand \Doi [0]{\begingroup \@sanitize@url \@Doi }%
\providecommand \@Doi  [1]{\endgroup\@@startlink{\doibase#1}\@@Doi}%
\providecommand \@@Doi [1]{#1\@@endlink}%
\providecommand \selectlanguage [0]{\@gobble}%
\providecommand \bibinfo  [0]{\@secondoftwo}%
\providecommand \bibfield  [0]{\@secondoftwo}%
\providecommand \translation [1]{[#1]}%
\providecommand \BibitemOpen [0]{}%
\providecommand \bibitemStop [0]{}%
\providecommand \bibitemNoStop [0]{.\EOS\space}%
\providecommand \EOS [0]{\spacefactor3000\relax}%
\providecommand \BibitemShut  [1]{\csname bibitem#1\endcsname}%
\bibitem [{\citenamefont {Adler}\ and\ \citenamefont
  {Davis}(1984)}]{Adler:1984ri}%
  \BibitemOpen
  \bibfield  {author} {\bibinfo {author} {\bibfnamefont {S.~L.}\ \bibnamefont
  {Adler}}\ and\ \bibinfo {author} {\bibfnamefont {A.}~\bibnamefont {Davis}},\
  }\Doi {10.1016/0550-3213(84)90324-9} {\bibfield  {journal} {\bibinfo
  {journal} {Nucl.Phys.},\ }\textbf {\bibinfo {volume} {B244}},\ \bibinfo
  {pages} {469} (\bibinfo {year} {1984})}\BibitemShut {NoStop}%
\bibitem [{\citenamefont {Alkofer}\ and\ \citenamefont
  {Amundsen}(1988)}]{Alkofer:1988tc}%
  \BibitemOpen
  \bibfield  {author} {\bibinfo {author} {\bibfnamefont {R.}~\bibnamefont
  {Alkofer}}\ and\ \bibinfo {author} {\bibfnamefont {P.}~\bibnamefont
  {Amundsen}},\ }\Doi {10.1016/0550-3213(88)90695-5} {\bibfield  {journal}
  {\bibinfo  {journal} {Nucl.Phys.},\ }\textbf {\bibinfo {volume} {B306}},\
  \bibinfo {pages} {305} (\bibinfo {year} {1988})}\BibitemShut {NoStop}%
\bibitem [{\citenamefont {Wagenbrunn}\ and\ \citenamefont
  {Glozman}(2007)}]{Wagenbrunn:2007ie}%
  \BibitemOpen
  \bibfield  {author} {\bibinfo {author} {\bibfnamefont {R.}~\bibnamefont
  {Wagenbrunn}}\ and\ \bibinfo {author} {\bibfnamefont {L.~Y.}\ \bibnamefont
  {Glozman}},\ }\Doi {10.1103/PhysRevD.75.036007} {\bibfield  {journal}
  {\bibinfo  {journal} {Phys.Rev.},\ }\textbf {\bibinfo {volume} {D75}},\
  \bibinfo {pages} {036007} (\bibinfo {year} {2007})},\ \Eprint
  {http://arxiv.org/abs/hep-ph/0701039} {arXiv:hep-ph/0701039 [hep-ph]}
  \BibitemShut {NoStop}%
\bibitem [{\citenamefont {Burgio}\ \emph
  {et~al.}(2012){\natexlab{a}}\citenamefont {Burgio}, \citenamefont
  {Schr{\"o}ck}, \citenamefont {Reinhardt},\ and\ \citenamefont
  {Quandt}}]{Burgio:2012ph}%
  \BibitemOpen
  \bibfield  {author} {\bibinfo {author} {\bibfnamefont {G.}~\bibnamefont
  {Burgio}}, \bibinfo {author} {\bibfnamefont {M.}~\bibnamefont {Schr{\"o}ck}},
  \bibinfo {author} {\bibfnamefont {H.}~\bibnamefont {Reinhardt}}, \ and\
  \bibinfo {author} {\bibfnamefont {M.}~\bibnamefont {Quandt}},\ }\Doi
  {10.1103/PhysRevD.86.014506} {\bibfield  {journal} {\bibinfo  {journal}
  {Phys.Rev.},\ }\textbf {\bibinfo {volume} {D86}},\ \bibinfo {pages} {014506}
  (\bibinfo {year} {2012}{\natexlab{a}})},\ \Eprint
  {http://arxiv.org/abs/1204.0716} {arXiv:1204.0716 [hep-lat]} \BibitemShut
  {NoStop}%
\bibitem [{\citenamefont {Burgio}\ \emph
  {et~al.}(2012){\natexlab{b}}\citenamefont {Burgio}, \citenamefont {Quandt},
  \citenamefont {Reinhardt},\ and\ \citenamefont
  {Schr{\"o}ck}}]{Burgio:2013mx}%
  \BibitemOpen
  \bibfield  {author} {\bibinfo {author} {\bibfnamefont {G.}~\bibnamefont
  {Burgio}}, \bibinfo {author} {\bibfnamefont {M.}~\bibnamefont {Quandt}},
  \bibinfo {author} {\bibfnamefont {H.}~\bibnamefont {Reinhardt}}, \ and\
  \bibinfo {author} {\bibfnamefont {M.}~\bibnamefont {Schr{\"o}ck}},\
  }\href@noop {} {\bibfield  {journal} {\bibinfo  {journal} {PoS},\ }\textbf
  {\bibinfo {volume} {ConfinementX}},\ \bibinfo {pages} {075} (\bibinfo {year}
  {2012}{\natexlab{b}})},\ \Eprint {http://arxiv.org/abs/1301.3619}
  {arXiv:1301.3619 [hep-lat]} \BibitemShut {NoStop}%
\bibitem [{\citenamefont {Bonnet}\ \emph {et~al.}(2002)\citenamefont {Bonnet},
  \citenamefont {Bowman}, \citenamefont {Leinweber}, \citenamefont {Williams},\
  and\ \citenamefont {Zhang}}]{Bonnet:2002ih}%
  \BibitemOpen
  \bibfield  {author} {\bibinfo {author} {\bibfnamefont {F.~D.}\ \bibnamefont
  {Bonnet}}, \bibinfo {author} {\bibfnamefont {P.~O.}\ \bibnamefont {Bowman}},
  \bibinfo {author} {\bibfnamefont {D.~B.}\ \bibnamefont {Leinweber}}, \bibinfo
  {author} {\bibfnamefont {A.~G.}\ \bibnamefont {Williams}}, \ and\ \bibinfo
  {author} {\bibfnamefont {J.-b.}\ \bibnamefont {Zhang}} (\bibinfo
  {collaboration} {CSSM Lattice collaboration}),\ }\Doi
  {10.1103/PhysRevD.65.114503} {\bibfield  {journal} {\bibinfo  {journal}
  {Phys.Rev.},\ }\textbf {\bibinfo {volume} {D65}},\ \bibinfo {pages} {114503}
  (\bibinfo {year} {2002})},\ \Eprint {http://arxiv.org/abs/hep-lat/0202003}
  {arXiv:hep-lat/0202003 [hep-lat]} \BibitemShut {NoStop}%
\bibitem [{\citenamefont {Zhang}\ \emph {et~al.}(2004)\citenamefont {Zhang},
  \citenamefont {Bowman}, \citenamefont {Leinweber}, \citenamefont {Williams},\
  and\ \citenamefont {Bonnet}}]{Zhang:2003faa}%
  \BibitemOpen
  \bibfield  {author} {\bibinfo {author} {\bibfnamefont {J.}~\bibnamefont
  {Zhang}}, \bibinfo {author} {\bibfnamefont {P.~O.}\ \bibnamefont {Bowman}},
  \bibinfo {author} {\bibfnamefont {D.~B.}\ \bibnamefont {Leinweber}}, \bibinfo
  {author} {\bibfnamefont {A.~G.}\ \bibnamefont {Williams}}, \ and\ \bibinfo
  {author} {\bibfnamefont {F.~D.}\ \bibnamefont {Bonnet}} (\bibinfo
  {collaboration} {CSSM Lattice collaboration}),\ }\Doi
  {10.1103/PhysRevD.70.034505} {\bibfield  {journal} {\bibinfo  {journal}
  {Phys.Rev.},\ }\textbf {\bibinfo {volume} {D70}},\ \bibinfo {pages} {034505}
  (\bibinfo {year} {2004})},\ \Eprint {http://arxiv.org/abs/hep-lat/0301018}
  {arXiv:hep-lat/0301018 [hep-lat]} \BibitemShut {NoStop}%
\bibitem [{\citenamefont {Zhang}\ \emph {et~al.}(2005)\citenamefont {Zhang},
  \citenamefont {Bowman}, \citenamefont {Coad}, \citenamefont {Heller},
  \citenamefont {Leinweber} \emph {et~al.}}]{Zhang:2004gv}%
  \BibitemOpen
  \bibfield  {author} {\bibinfo {author} {\bibfnamefont {J.}~\bibnamefont
  {Zhang}}, \bibinfo {author} {\bibfnamefont {P.~O.}\ \bibnamefont {Bowman}},
  \bibinfo {author} {\bibfnamefont {R.~J.}\ \bibnamefont {Coad}}, \bibinfo
  {author} {\bibfnamefont {U.~M.}\ \bibnamefont {Heller}}, \bibinfo {author}
  {\bibfnamefont {D.~B.}\ \bibnamefont {Leinweber}},  \emph {et~al.},\ }\Doi
  {10.1103/PhysRevD.71.014501} {\bibfield  {journal} {\bibinfo  {journal}
  {Phys.Rev.},\ }\textbf {\bibinfo {volume} {D71}},\ \bibinfo {pages} {014501}
  (\bibinfo {year} {2005})},\ \Eprint {http://arxiv.org/abs/hep-lat/0410045}
  {arXiv:hep-lat/0410045 [hep-lat]} \BibitemShut {NoStop}%
\bibitem [{\citenamefont {Trewartha}\ \emph {et~al.}(2013)\citenamefont
  {Trewartha}, \citenamefont {Kamleh}, \citenamefont {Leinweber},\ and\
  \citenamefont {Roberts}}]{Trewartha:2013qga}%
  \BibitemOpen
  \bibfield  {author} {\bibinfo {author} {\bibfnamefont {D.}~\bibnamefont
  {Trewartha}}, \bibinfo {author} {\bibfnamefont {W.}~\bibnamefont {Kamleh}},
  \bibinfo {author} {\bibfnamefont {D.}~\bibnamefont {Leinweber}}, \ and\
  \bibinfo {author} {\bibfnamefont {D.~S.}\ \bibnamefont {Roberts}},\ }\Doi
  {10.1103/PhysRevD.88.034501} {\bibfield  {journal} {\bibinfo  {journal}
  {Phys.Rev.},\ }\textbf {\bibinfo {volume} {D88}},\ \bibinfo {pages} {034501}
  (\bibinfo {year} {2013})},\ \Eprint {http://arxiv.org/abs/1306.3283}
  {arXiv:1306.3283 [hep-lat]} \BibitemShut {NoStop}%
\bibitem [{\citenamefont {Guo}\ and\ \citenamefont
  {Szczepaniak}(2009)}]{Guo:2009ma}%
  \BibitemOpen
  \bibfield  {author} {\bibinfo {author} {\bibfnamefont {P.}~\bibnamefont
  {Guo}}\ and\ \bibinfo {author} {\bibfnamefont {A.~P.}\ \bibnamefont
  {Szczepaniak}},\ }\Doi {10.1103/PhysRevD.79.116006} {\bibfield  {journal}
  {\bibinfo  {journal} {Phys.Rev.},\ }\textbf {\bibinfo {volume} {D79}},\
  \bibinfo {pages} {116006} (\bibinfo {year} {2009})},\ \Eprint
  {http://arxiv.org/abs/0902.1316} {arXiv:0902.1316 [hep-ph]} \BibitemShut
  {NoStop}%
\bibitem [{\citenamefont {Pak}\ and\ \citenamefont
  {Glozman}(2013)}]{Pak:2013cpa}%
  \BibitemOpen
  \bibfield  {author} {\bibinfo {author} {\bibfnamefont {M.}~\bibnamefont
  {Pak}}\ and\ \bibinfo {author} {\bibfnamefont {L.~Y.}\ \bibnamefont
  {Glozman}},\ }\Doi {10.1103/PhysRevD.88.076010} {\bibfield  {journal}
  {\bibinfo  {journal} {Phys.Rev.},\ }\textbf {\bibinfo {volume} {D88}},\
  \bibinfo {pages} {076010} (\bibinfo {year} {2013})},\ \Eprint
  {http://arxiv.org/abs/1307.4224} {arXiv:1307.4224 [hep-ph]} \BibitemShut
  {NoStop}%
\bibitem [{\citenamefont {Pak}\ and\ \citenamefont
  {Reinhardt}(2012)}]{Pak:2011wu}%
  \BibitemOpen
  \bibfield  {author} {\bibinfo {author} {\bibfnamefont {M.}~\bibnamefont
  {Pak}}\ and\ \bibinfo {author} {\bibfnamefont {H.}~\bibnamefont
  {Reinhardt}},\ }\Doi {10.1016/j.physletb.2012.01.006} {\bibfield  {journal}
  {\bibinfo  {journal} {Phys.Lett.},\ }\textbf {\bibinfo {volume} {B707}},\
  \bibinfo {pages} {566} (\bibinfo {year} {2012})},\ \Eprint
  {http://arxiv.org/abs/1107.5263} {arXiv:1107.5263 [hep-ph]} \BibitemShut
  {NoStop}%
\bibitem [{\citenamefont {Pak}\ and\ \citenamefont
  {Reinhardt}(2013)}]{Pak:2013uba}%
  \BibitemOpen
  \bibfield  {author} {\bibinfo {author} {\bibfnamefont {M.}~\bibnamefont
  {Pak}}\ and\ \bibinfo {author} {\bibfnamefont {H.}~\bibnamefont
  {Reinhardt}},\ }\Doi {10.1103/PhysRevD.88.125021} {\bibfield  {journal}
  {\bibinfo  {journal} {Phys.Rev.},\ }\textbf {\bibinfo {volume} {D88}},\
  \bibinfo {pages} {125021} (\bibinfo {year} {2013})},\ \Eprint
  {http://arxiv.org/abs/1310.1797} {arXiv:1310.1797 [hep-ph]} \BibitemShut
  {NoStop}%
\bibitem [{\citenamefont {Lang}\ and\ \citenamefont
  {Schr{\"o}ck}(2011)}]{Lang:2011vw}%
  \BibitemOpen
  \bibfield  {author} {\bibinfo {author} {\bibfnamefont {C.~B.}\ \bibnamefont
  {Lang}}\ and\ \bibinfo {author} {\bibfnamefont {M.}~\bibnamefont
  {Schr{\"o}ck}},\ }\Doi {10.1103/PhysRevD.84.087704} {\bibfield  {journal}
  {\bibinfo  {journal} {Phys.Rev.},\ }\textbf {\bibinfo {volume} {D84}},\
  \bibinfo {pages} {087704} (\bibinfo {year} {2011})},\ \Eprint
  {http://arxiv.org/abs/1107.5195} {arXiv:1107.5195 [hep-lat]} \BibitemShut
  {NoStop}%
\bibitem [{\citenamefont {Glozman}\ \emph {et~al.}(2012)\citenamefont
  {Glozman}, \citenamefont {Lang},\ and\ \citenamefont
  {Schr{\"o}ck}}]{Glozman:2012fj}%
  \BibitemOpen
  \bibfield  {author} {\bibinfo {author} {\bibfnamefont {L.~Y.}\ \bibnamefont
  {Glozman}}, \bibinfo {author} {\bibfnamefont {C.~B.}\ \bibnamefont {Lang}}, \
  and\ \bibinfo {author} {\bibfnamefont {M.}~\bibnamefont {Schr{\"o}ck}},\
  }\Doi {10.1103/PhysRevD.86.014507} {\bibfield  {journal} {\bibinfo  {journal}
  {Phys.Rev.},\ }\textbf {\bibinfo {volume} {D86}},\ \bibinfo {pages} {014507}
  (\bibinfo {year} {2012})},\ \Eprint {http://arxiv.org/abs/1205.4887}
  {arXiv:1205.4887 [hep-lat]} \BibitemShut {NoStop}%
\bibitem [{\citenamefont {Denissenya}\ \emph {et~al.}(2014)\citenamefont
  {Denissenya}, \citenamefont {Glozman},\ and\ \citenamefont
  {Lang}}]{Denissenya:2014poa}%
  \BibitemOpen
  \bibfield  {author} {\bibinfo {author} {\bibfnamefont {M.}~\bibnamefont
  {Denissenya}}, \bibinfo {author} {\bibfnamefont {L.~Y.}\ \bibnamefont
  {Glozman}}, \ and\ \bibinfo {author} {\bibfnamefont {C.~B.}\ \bibnamefont
  {Lang}},\ }\Doi {10.1103/PhysRevD.89.077502} {\bibfield  {journal} {\bibinfo
  {journal} {Phys.Rev.},\ }\textbf {\bibinfo {volume} {D89}},\ \bibinfo {pages}
  {077502} (\bibinfo {year} {2014})},\ \Eprint {http://arxiv.org/abs/1402.1887}
  {arXiv:1402.1887 [hep-lat]} \BibitemShut {NoStop}%
\bibitem [{\citenamefont {Schr{\"o}ck}(2012)}]{Schrock:2011hq}%
  \BibitemOpen
  \bibfield  {author} {\bibinfo {author} {\bibfnamefont {M.}~\bibnamefont
  {Schr{\"o}ck}},\ }\Doi {10.1016/j.physletb.2012.04.008} {\bibfield  {journal}
  {\bibinfo  {journal} {Phys.Lett.},\ }\textbf {\bibinfo {volume} {B711}},\
  \bibinfo {pages} {217} (\bibinfo {year} {2012})},\ \Eprint
  {http://arxiv.org/abs/1112.5107} {arXiv:1112.5107 [hep-lat]} \BibitemShut
  {NoStop}%
\bibitem [{\citenamefont {Neuberger}(1998){\natexlab{a}}}]{Neuberger:1997fp}%
  \BibitemOpen
  \bibfield  {author} {\bibinfo {author} {\bibfnamefont {H.}~\bibnamefont
  {Neuberger}},\ }\Doi {10.1016/S0370-2693(97)01368-3} {\bibfield  {journal}
  {\bibinfo  {journal} {Phys.Lett.},\ }\textbf {\bibinfo {volume} {B417}},\
  \bibinfo {pages} {141} (\bibinfo {year} {1998}{\natexlab{a}})},\ \Eprint
  {http://arxiv.org/abs/hep-lat/9707022} {arXiv:hep-lat/9707022 [hep-lat]}
  \BibitemShut {NoStop}%
\bibitem [{\citenamefont {Neuberger}(1998){\natexlab{b}}}]{Neuberger:1998wv}%
  \BibitemOpen
  \bibfield  {author} {\bibinfo {author} {\bibfnamefont {H.}~\bibnamefont
  {Neuberger}},\ }\Doi {10.1016/S0370-2693(98)00355-4} {\bibfield  {journal}
  {\bibinfo  {journal} {Phys.Lett.},\ }\textbf {\bibinfo {volume} {B427}},\
  \bibinfo {pages} {353} (\bibinfo {year} {1998}{\natexlab{b}})},\ \Eprint
  {http://arxiv.org/abs/hep-lat/9801031} {arXiv:hep-lat/9801031 [hep-lat]}
  \BibitemShut {NoStop}%
\bibitem [{\citenamefont {Ginsparg}\ and\ \citenamefont
  {Wilson}(1982)}]{Ginsparg:1981bj}%
  \BibitemOpen
  \bibfield  {author} {\bibinfo {author} {\bibfnamefont {P.~H.}\ \bibnamefont
  {Ginsparg}}\ and\ \bibinfo {author} {\bibfnamefont {K.~G.}\ \bibnamefont
  {Wilson}},\ }\href@noop {} {\bibfield  {journal} {\bibinfo  {journal}
  {Phys.Rev.},\ }\textbf {\bibinfo {volume} {D25}},\ \bibinfo {pages} {2649}
  (\bibinfo {year} {1982})}\BibitemShut {NoStop}%
\bibitem [{\citenamefont {Hasenfratz}\ \emph {et~al.}(1998)\citenamefont
  {Hasenfratz}, \citenamefont {Laliena},\ and\ \citenamefont
  {Niedermayer}}]{Hasenfratz:1998ri}%
  \BibitemOpen
  \bibfield  {author} {\bibinfo {author} {\bibfnamefont {P.}~\bibnamefont
  {Hasenfratz}}, \bibinfo {author} {\bibfnamefont {V.}~\bibnamefont {Laliena}},
  \ and\ \bibinfo {author} {\bibfnamefont {F.}~\bibnamefont {Niedermayer}},\
  }\Doi {10.1016/S0370-2693(98)00315-3} {\bibfield  {journal} {\bibinfo
  {journal} {Phys.Lett.},\ }\textbf {\bibinfo {volume} {B427}},\ \bibinfo
  {pages} {125} (\bibinfo {year} {1998})},\ \Eprint
  {http://arxiv.org/abs/hep-lat/9801021} {arXiv:hep-lat/9801021 [hep-lat]}
  \BibitemShut {NoStop}%
\bibitem [{\citenamefont {Hasenfratz}(1998)}]{Hasenfratz:1998jp}%
  \BibitemOpen
  \bibfield  {author} {\bibinfo {author} {\bibfnamefont {P.}~\bibnamefont
  {Hasenfratz}},\ }\Doi {10.1016/S0550-3213(98)00399-X} {\bibfield  {journal}
  {\bibinfo  {journal} {Nucl.Phys.},\ }\textbf {\bibinfo {volume} {B525}},\
  \bibinfo {pages} {401} (\bibinfo {year} {1998})},\ \Eprint
  {http://arxiv.org/abs/hep-lat/9802007} {arXiv:hep-lat/9802007 [hep-lat]}
  \BibitemShut {NoStop}%
\bibitem [{\citenamefont {Luscher}(1998)}]{Luscher:1998pqa}%
  \BibitemOpen
  \bibfield  {author} {\bibinfo {author} {\bibfnamefont {M.}~\bibnamefont
  {Luscher}},\ }\Doi {10.1016/S0370-2693(98)00423-7} {\bibfield  {journal}
  {\bibinfo  {journal} {Phys.Lett.},\ }\textbf {\bibinfo {volume} {B428}},\
  \bibinfo {pages} {342} (\bibinfo {year} {1998})},\ \Eprint
  {http://arxiv.org/abs/hep-lat/9802011} {arXiv:hep-lat/9802011 [hep-lat]}
  \BibitemShut {NoStop}%
\bibitem [{\citenamefont {Skullerud}\ and\ \citenamefont
  {Williams}(2001)}]{Skullerud:2000un}%
  \BibitemOpen
  \bibfield  {author} {\bibinfo {author} {\bibfnamefont {J.~I.}\ \bibnamefont
  {Skullerud}}\ and\ \bibinfo {author} {\bibfnamefont {A.~G.}\ \bibnamefont
  {Williams}},\ }\Doi {10.1103/PhysRevD.63.054508} {\bibfield  {journal}
  {\bibinfo  {journal} {Phys.Rev.},\ }\textbf {\bibinfo {volume} {D63}},\
  \bibinfo {pages} {054508} (\bibinfo {year} {2001})},\ \Eprint
  {http://arxiv.org/abs/hep-lat/0007028} {arXiv:hep-lat/0007028 [hep-lat]}
  \BibitemShut {NoStop}%
\bibitem [{\citenamefont {L{\"u}scher}\ and\ \citenamefont
  {Weisz}(1985)}]{Luscher:1984xn}%
  \BibitemOpen
  \bibfield  {author} {\bibinfo {author} {\bibfnamefont {M.}~\bibnamefont
  {L{\"u}scher}}\ and\ \bibinfo {author} {\bibfnamefont {P.}~\bibnamefont
  {Weisz}},\ }\Doi {10.1007/BF01206178} {\bibfield  {journal} {\bibinfo
  {journal} {Commun.Math.Phys.},\ }\textbf {\bibinfo {volume} {97}},\ \bibinfo
  {pages} {59} (\bibinfo {year} {1985})}\BibitemShut {NoStop}%
\bibitem [{\citenamefont {Edwards}\ and\ \citenamefont
  {Joo}(2005)}]{Edwards:2004sx}%
  \BibitemOpen
  \bibfield  {author} {\bibinfo {author} {\bibfnamefont {R.~G.}\ \bibnamefont
  {Edwards}}\ and\ \bibinfo {author} {\bibfnamefont {B.}~\bibnamefont {Joo}}
  (\bibinfo {collaboration} {SciDAC Collaboration, LHPC Collaboration, UKQCD
  Collaboration}),\ }\Doi {10.1016/j.nuclphysbps.2004.11.254} {\bibfield
  {journal} {\bibinfo  {journal} {Nucl.Phys.Proc.Suppl.},\ }\textbf {\bibinfo
  {volume} {140}},\ \bibinfo {pages} {832} (\bibinfo {year} {2005})},\ \Eprint
  {http://arxiv.org/abs/hep-lat/0409003} {arXiv:hep-lat/0409003 [hep-lat]}
  \BibitemShut {NoStop}%
\bibitem [{\citenamefont {Winter}(2013)}]{Winter:2014npa}%
  \BibitemOpen
  \bibfield  {author} {\bibinfo {author} {\bibfnamefont {F.}~\bibnamefont
  {Winter}},\ }\href@noop {} {\bibfield  {journal} {\bibinfo  {journal} {PoS},\
  }\textbf {\bibinfo {volume} {LATTICE2013}},\ \bibinfo {pages} {042} (\bibinfo
  {year} {2013})}\BibitemShut {NoStop}%
\bibitem [{\citenamefont {Winter}\ \emph {et~al.}(2014)\citenamefont {Winter},
  \citenamefont {Clark}, \citenamefont {Edwards},\ and\ \citenamefont
  {Joó}}]{Winter:2014dka}%
  \BibitemOpen
  \bibfield  {author} {\bibinfo {author} {\bibfnamefont {F.}~\bibnamefont
  {Winter}}, \bibinfo {author} {\bibfnamefont {M.}~\bibnamefont {Clark}},
  \bibinfo {author} {\bibfnamefont {R.}~\bibnamefont {Edwards}}, \ and\
  \bibinfo {author} {\bibfnamefont {B.}~\bibnamefont {Joó}},\ }\Doi
  {10.1109/IPDPS.2014.112} { (\bibinfo {year} {2014})},\ \doi
  {10.1109/IPDPS.2014.112},\ \Eprint {http://arxiv.org/abs/1408.5925}
  {arXiv:1408.5925 [hep-lat]} \BibitemShut {NoStop}%
\bibitem [{\citenamefont {Gattringer}\ \emph {et~al.}(2002)\citenamefont
  {Gattringer}, \citenamefont {Hoffmann},\ and\ \citenamefont
  {Schaefer}}]{Gattringer:2001jf}%
  \BibitemOpen
  \bibfield  {author} {\bibinfo {author} {\bibfnamefont {C.}~\bibnamefont
  {Gattringer}}, \bibinfo {author} {\bibfnamefont {R.}~\bibnamefont
  {Hoffmann}}, \ and\ \bibinfo {author} {\bibfnamefont {S.}~\bibnamefont
  {Schaefer}},\ }\Doi {10.1103/PhysRevD.65.094503} {\bibfield  {journal}
  {\bibinfo  {journal} {Phys.Rev.},\ }\textbf {\bibinfo {volume} {D65}},\
  \bibinfo {pages} {094503} (\bibinfo {year} {2002})},\ \Eprint
  {http://arxiv.org/abs/hep-lat/0112024} {arXiv:hep-lat/0112024 [hep-lat]}
  \BibitemShut {NoStop}%
\bibitem [{\citenamefont {Skullerud}\ and\ \citenamefont
  {Williams}(2000)}]{Skullerud:1999gv}%
  \BibitemOpen
  \bibfield  {author} {\bibinfo {author} {\bibfnamefont {J.~I.}\ \bibnamefont
  {Skullerud}}\ and\ \bibinfo {author} {\bibfnamefont {A.~G.}\ \bibnamefont
  {Williams}},\ }\href@noop {} {\bibfield  {journal} {\bibinfo  {journal}
  {Nucl.Phys.Proc.Suppl.},\ }\textbf {\bibinfo {volume} {83}},\ \bibinfo
  {pages} {209} (\bibinfo {year} {2000})},\ \Eprint
  {http://arxiv.org/abs/hep-lat/9909142} {arXiv:hep-lat/9909142 [hep-lat]}
  \BibitemShut {NoStop}%
\bibitem [{\citenamefont {Schr{\"o}ck}\ and\ \citenamefont
  {Vogt}(2013)}]{Schrock:2012fj}%
  \BibitemOpen
  \bibfield  {author} {\bibinfo {author} {\bibfnamefont {M.}~\bibnamefont
  {Schr{\"o}ck}}\ and\ \bibinfo {author} {\bibfnamefont {H.}~\bibnamefont
  {Vogt}},\ }\Doi {10.1016/j.cpc.2013.03.021} {\bibfield  {journal} {\bibinfo
  {journal} {Comput.Phys.Commun.},\ }\textbf {\bibinfo {volume} {184}},\
  \bibinfo {pages} {1907} (\bibinfo {year} {2013})},\ \Eprint
  {http://arxiv.org/abs/1212.5221} {arXiv:1212.5221 [hep-lat]} \BibitemShut
  {NoStop}%
\bibitem [{\citenamefont {Mandula}\ and\ \citenamefont
  {Ogilvie}(1990)}]{Mandula:1990vs}%
  \BibitemOpen
  \bibfield  {author} {\bibinfo {author} {\bibfnamefont {J.~E.}\ \bibnamefont
  {Mandula}}\ and\ \bibinfo {author} {\bibfnamefont {M.}~\bibnamefont
  {Ogilvie}},\ }\Doi {10.1016/0370-2693(90)90031-Z} {\bibfield  {journal}
  {\bibinfo  {journal} {Phys.Lett.},\ }\textbf {\bibinfo {volume} {B248}},\
  \bibinfo {pages} {156} (\bibinfo {year} {1990})}\BibitemShut {NoStop}%
\bibitem [{\citenamefont {Burgio}\ \emph {et~al.}(2009)\citenamefont {Burgio},
  \citenamefont {Quandt},\ and\ \citenamefont {Reinhardt}}]{Burgio:2008jr}%
  \BibitemOpen
  \bibfield  {author} {\bibinfo {author} {\bibfnamefont {G.}~\bibnamefont
  {Burgio}}, \bibinfo {author} {\bibfnamefont {M.}~\bibnamefont {Quandt}}, \
  and\ \bibinfo {author} {\bibfnamefont {H.}~\bibnamefont {Reinhardt}},\ }\Doi
  {10.1103/PhysRevLett.102.032002} {\bibfield  {journal} {\bibinfo  {journal}
  {Phys.Rev.Lett.},\ }\textbf {\bibinfo {volume} {102}},\ \bibinfo {pages}
  {032002} (\bibinfo {year} {2009})},\ \Eprint {http://arxiv.org/abs/0807.3291}
  {arXiv:0807.3291 [hep-lat]} \BibitemShut {NoStop}%
\bibitem [{\citenamefont {Gattringer}\ and\ \citenamefont
  {Lang}(2010)}]{Gattringer:2010zz}%
  \BibitemOpen
  \bibfield  {author} {\bibinfo {author} {\bibfnamefont {C.}~\bibnamefont
  {Gattringer}}\ and\ \bibinfo {author} {\bibfnamefont {C.~B.}\ \bibnamefont
  {Lang}},\ }\Doi {10.1007/978-3-642-01850-3} {\bibfield  {journal} {\bibinfo
  {journal} {Lect.Notes Phys.},\ }\textbf {\bibinfo {volume} {788}},\ \bibinfo
  {pages} {1} (\bibinfo {year} {2010})}\BibitemShut {NoStop}%
\bibitem [{\citenamefont {Lehoucq}\ \emph {et~al.}(1998)\citenamefont
  {Lehoucq}, \citenamefont {Sorensen},\ and\ \citenamefont {Yang}}]{Arpack}%
  \BibitemOpen
  \bibfield  {author} {\bibinfo {author} {\bibfnamefont {R.~B.}\ \bibnamefont
  {Lehoucq}}, \bibinfo {author} {\bibfnamefont {D.~C.}\ \bibnamefont
  {Sorensen}}, \ and\ \bibinfo {author} {\bibfnamefont {C.}~\bibnamefont
  {Yang}},\ }\href@noop {} {\bibfield  {journal} {\bibinfo  {journal} {SIAM,
  New Vork}} (\bibinfo {year} {1998})}\BibitemShut {NoStop}%
\bibitem [{\citenamefont {Jegerlehner}(1996)}]{Jegerlehner:1996pm}%
  \BibitemOpen
  \bibfield  {author} {\bibinfo {author} {\bibfnamefont {B.}~\bibnamefont
  {Jegerlehner}},\ }\href@noop {} { (\bibinfo {year} {1996})},\ \Eprint
  {http://arxiv.org/abs/hep-lat/9612014} {arXiv:hep-lat/9612014 [hep-lat]}
  \BibitemShut {NoStop}%
\bibitem [{\citenamefont {Morningstar}\ and\ \citenamefont
  {Peardon}(2004)}]{Morningstar:2003gk}%
  \BibitemOpen
  \bibfield  {author} {\bibinfo {author} {\bibfnamefont {C.}~\bibnamefont
  {Morningstar}}\ and\ \bibinfo {author} {\bibfnamefont {M.~J.}\ \bibnamefont
  {Peardon}},\ }\Doi {10.1103/PhysRevD.69.054501} {\bibfield  {journal}
  {\bibinfo  {journal} {Phys.Rev.},\ }\textbf {\bibinfo {volume} {D69}},\
  \bibinfo {pages} {054501} (\bibinfo {year} {2004})},\ \Eprint
  {http://arxiv.org/abs/hep-lat/0311018} {arXiv:hep-lat/0311018 [hep-lat]}
  \BibitemShut {NoStop}%
\bibitem [{\citenamefont {Neuberger}(2000)}]{Neuberger:1999pz}%
  \BibitemOpen
  \bibfield  {author} {\bibinfo {author} {\bibfnamefont {H.}~\bibnamefont
  {Neuberger}},\ }\Doi {10.1103/PhysRevD.61.085015} {\bibfield  {journal}
  {\bibinfo  {journal} {Phys.Rev.},\ }\textbf {\bibinfo {volume} {D61}},\
  \bibinfo {pages} {085015} (\bibinfo {year} {2000})},\ \Eprint
  {http://arxiv.org/abs/hep-lat/9911004} {arXiv:hep-lat/9911004 [hep-lat]}
  \BibitemShut {NoStop}%
\bibitem [{\citenamefont {Popovici}\ \emph {et~al.}(2009)\citenamefont
  {Popovici}, \citenamefont {Watson},\ and\ \citenamefont
  {Reinhardt}}]{Popovici:2008ty}%
  \BibitemOpen
  \bibfield  {author} {\bibinfo {author} {\bibfnamefont {C.}~\bibnamefont
  {Popovici}}, \bibinfo {author} {\bibfnamefont {P.}~\bibnamefont {Watson}}, \
  and\ \bibinfo {author} {\bibfnamefont {H.}~\bibnamefont {Reinhardt}},\ }\Doi
  {10.1103/PhysRevD.79.045006} {\bibfield  {journal} {\bibinfo  {journal}
  {Phys.Rev.},\ }\textbf {\bibinfo {volume} {D79}},\ \bibinfo {pages} {045006}
  (\bibinfo {year} {2009})},\ \Eprint {http://arxiv.org/abs/0810.4887}
  {arXiv:0810.4887 [hep-th]} \BibitemShut {NoStop}%
\bibitem [{\citenamefont {Watson}\ and\ \citenamefont
  {Reinhardt}(2012){\natexlab{a}}}]{Watson:2011kv}%
  \BibitemOpen
  \bibfield  {author} {\bibinfo {author} {\bibfnamefont {P.}~\bibnamefont
  {Watson}}\ and\ \bibinfo {author} {\bibfnamefont {H.}~\bibnamefont
  {Reinhardt}},\ }\Doi {10.1103/PhysRevD.85.025014} {\bibfield  {journal}
  {\bibinfo  {journal} {Phys.Rev.},\ }\textbf {\bibinfo {volume} {D85}},\
  \bibinfo {pages} {025014} (\bibinfo {year} {2012}{\natexlab{a}})},\ \Eprint
  {http://arxiv.org/abs/1111.6078} {arXiv:1111.6078 [hep-ph]} \BibitemShut
  {NoStop}%
\bibitem [{\citenamefont {Watson}\ and\ \citenamefont
  {Reinhardt}(2012){\natexlab{b}}}]{Watson:2012ht}%
  \BibitemOpen
  \bibfield  {author} {\bibinfo {author} {\bibfnamefont {P.}~\bibnamefont
  {Watson}}\ and\ \bibinfo {author} {\bibfnamefont {H.}~\bibnamefont
  {Reinhardt}},\ }\Doi {10.1103/PhysRevD.86.125030} {\bibfield  {journal}
  {\bibinfo  {journal} {Phys.Rev.},\ }\textbf {\bibinfo {volume} {D86}},\
  \bibinfo {pages} {125030} (\bibinfo {year} {2012}{\natexlab{b}})},\ \Eprint
  {http://arxiv.org/abs/1211.4507} {arXiv:1211.4507 [hep-ph]} \BibitemShut
  {NoStop}%
\bibitem [{\citenamefont {Zhang}\ \emph {et~al.}(2009)\citenamefont {Zhang},
  \citenamefont {Moran}, \citenamefont {Bowman}, \citenamefont {Leinweber},\
  and\ \citenamefont {Williams}}]{Zhang:2009jf}%
  \BibitemOpen
  \bibfield  {author} {\bibinfo {author} {\bibfnamefont {J.}~\bibnamefont
  {Zhang}}, \bibinfo {author} {\bibfnamefont {P.~J.}\ \bibnamefont {Moran}},
  \bibinfo {author} {\bibfnamefont {P.~O.}\ \bibnamefont {Bowman}}, \bibinfo
  {author} {\bibfnamefont {D.~B.}\ \bibnamefont {Leinweber}}, \ and\ \bibinfo
  {author} {\bibfnamefont {A.~G.}\ \bibnamefont {Williams}},\ }\Doi
  {10.1103/PhysRevD.80.074503} {\bibfield  {journal} {\bibinfo  {journal}
  {Phys.Rev.},\ }\textbf {\bibinfo {volume} {D80}},\ \bibinfo {pages} {074503}
  (\bibinfo {year} {2009})},\ \Eprint {http://arxiv.org/abs/0908.3726}
  {arXiv:0908.3726 [hep-lat]} \BibitemShut {NoStop}%
\bibitem [{\citenamefont {Banks}\ and\ \citenamefont
  {Casher}(1980)}]{Banks:1979yr}%
  \BibitemOpen
  \bibfield  {author} {\bibinfo {author} {\bibfnamefont {T.}~\bibnamefont
  {Banks}}\ and\ \bibinfo {author} {\bibfnamefont {A.}~\bibnamefont {Casher}},\
  }\Doi {10.1016/0550-3213(80)90255-2} {\bibfield  {journal} {\bibinfo
  {journal} {Nucl.Phys.},\ }\textbf {\bibinfo {volume} {B169}},\ \bibinfo
  {pages} {103} (\bibinfo {year} {1980})}\BibitemShut {NoStop}%
\bibitem [{\citenamefont {Schr{\"o}ck}\ \emph {et~al.}(2014)\citenamefont
  {Schr{\"o}ck}, \citenamefont {Denissenya}, \citenamefont {Glozman},\ and\
  \citenamefont {Lang}}]{Schrock:2013rea}%
  \BibitemOpen
  \bibfield  {author} {\bibinfo {author} {\bibfnamefont {M.}~\bibnamefont
  {Schr{\"o}ck}}, \bibinfo {author} {\bibfnamefont {M.}~\bibnamefont
  {Denissenya}}, \bibinfo {author} {\bibfnamefont {L.~Y.}\ \bibnamefont
  {Glozman}}, \ and\ \bibinfo {author} {\bibfnamefont {C.~B.}\ \bibnamefont
  {Lang}},\ }\href@noop {} {\bibfield  {journal} {\bibinfo  {journal} {PoS},\
  }\textbf {\bibinfo {volume} {LATTICE2013}},\ \bibinfo {pages} {116} (\bibinfo
  {year} {2014})},\ \Eprint {http://arxiv.org/abs/1309.0202} {arXiv:1309.0202
  [hep-lat]} \BibitemShut {NoStop}%
\bibitem [{\citenamefont {Hernandez}\ \emph {et~al.}(1999)\citenamefont
  {Hernandez}, \citenamefont {Jansen},\ and\ \citenamefont
  {Luscher}}]{Hernandez:1998et}%
  \BibitemOpen
  \bibfield  {author} {\bibinfo {author} {\bibfnamefont {P.}~\bibnamefont
  {Hernandez}}, \bibinfo {author} {\bibfnamefont {K.}~\bibnamefont {Jansen}}, \
  and\ \bibinfo {author} {\bibfnamefont {M.}~\bibnamefont {Luscher}},\ }\Doi
  {10.1016/S0550-3213(99)00213-8} {\bibfield  {journal} {\bibinfo  {journal}
  {Nucl.Phys.},\ }\textbf {\bibinfo {volume} {B552}},\ \bibinfo {pages} {363}
  (\bibinfo {year} {1999})},\ \Eprint {http://arxiv.org/abs/hep-lat/9808010}
  {arXiv:hep-lat/9808010 [hep-lat]} \BibitemShut {NoStop}%
\bibitem [{\citenamefont {Glozman}\ and\ \citenamefont
  {Wagenbrunn}(2008)}]{Glozman:2007tv}%
  \BibitemOpen
  \bibfield  {author} {\bibinfo {author} {\bibfnamefont {L.~Y.}\ \bibnamefont
  {Glozman}}\ and\ \bibinfo {author} {\bibfnamefont {R.}~\bibnamefont
  {Wagenbrunn}},\ }\Doi {10.1103/PhysRevD.77.054027} {\bibfield  {journal}
  {\bibinfo  {journal} {Phys.Rev.},\ }\textbf {\bibinfo {volume} {D77}},\
  \bibinfo {pages} {054027} (\bibinfo {year} {2008})},\ \Eprint
  {http://arxiv.org/abs/0709.3080} {arXiv:0709.3080 [hep-ph]} \BibitemShut
  {NoStop}%
\end{thebibliography}%

\end{document}